\documentclass[reprint,prb,amsmath,amssymb,aps]{revtex4-2}

\usepackage{graphicx}
\usepackage{dcolumn}
\usepackage{bm}
\usepackage{float}
\usepackage{xcolor}

\usepackage{feynmp-auto}    

\usepackage{tikz}  
\usetikzlibrary{decorations.pathmorphing,arrows.meta} 
\usetikzlibrary{decorations.markings}  

\begin{document}

\title{Impact of spin-orbit coupling on electron correlation corrections to the density of states in anisotropic conductors}

\author{Bahruz Suleymanli}
\email{bahruz.suleymanli@gmail.com}
\affiliation{Physics Department, Yıldız Technical University, 34220 Esenler, Istanbul, Türkiye}

\author{B. Tanatar}
\email{tanatar@fen.bilkent.edu.tr}
\affiliation{Department of Physics, Bilkent University, 06800 Ankara, Türkiye}

\date{\today}

\begin{abstract}
We study Altshuler–Aronov-type interaction corrections to the single-particle density of states 
(DOS) in a strongly anisotropic 2D conductor with an open Fermi surface (FS) and weak disorder, 
in the presence of coexisting Rashba and Dresselhaus spin–orbit couplings (SOCs) constrained to 
the longitudinal direction. The low-energy band consists of two warped sheets weakly tunnel-coupled 
transversely; SOC splits the sheets into helicity branches with a fixed spin axis. Working in 
a Matsubara space, we compute the exchange contribution in the diffusion channel with dynamically 
screened Coulomb interaction and an impurity ladder.
The resulting DOS anomaly exhibits a dimensional crossover governed by the transverse coupling scale $\varepsilon_c$. Close to the Fermi level ($|\varepsilon-\varepsilon_F|<\varepsilon_c$), the system behaves two-dimensionally, featuring a logarithmic DOS dip whose magnitude is enhanced by intrinsic SOCs. Further from the Fermi level ($|\varepsilon-\varepsilon_F|\!>\!\varepsilon_c$), the system behaves quasi-one-dimensionally, featuring a sharper square-root singularity whose amplitude is remarkably enhanced by the SOCs. Notably, we identify a critical SOC strength at which these spin-orbit effects exactly cancel the electron-correlation correction, perfectly restoring the unperturbed density of states. Furthermore, increasing the SOC beyond this critical point inverts the sign of the anomaly entirely, yielding a positive DOS correction.
This sign reversal fundamentally alters the energy dependence, such that at energies beyond $\varepsilon_F + \varepsilon_c$, the positive correction decays to smaller values as energy increases, opposite to the standard negative correction. 
This contrasting trend provides a distinct spectroscopic signature of SOC-modulated correlation effects.
\end{abstract}

\maketitle

\section{\label{sec:Introduction}Introduction}

Low-energy irregularities—most notably zero-bias anomalies (ZBAs)—are frequently observed in 
modern low-dimensional conductors, including gate-tunable two-dimensional electron gases 
(2DEGs) with Rashba spin–orbit coupling (SOC) \cite{Miller2003, Caviglia2010, Guo2021}, 2D 
materials with proximity-induced SOC, such as graphene/TMD heterostructures \cite{Mariani2007, 
Wang2015, Wakamura2018, Albin2021, Sun2023}, 2D electron systems patterned by 
unidirectional lateral superlattices \cite{Endo2005, Endo2021}, self-assembled quasi-1D 
surface atomic chains \cite{Blumenstein2011, Tegenkamp2012, Park2013}, and ultra-thin 
disordered metallic films \cite{Ossi2013, Ovadyahu2022}. However, in practice, distinct 
microscopic processes—Altshuler–Aronov (AA) interaction anomalies that suppress 
single-particle DOS in a dimensionality-dependent way \cite{Altshuler1979}, Coulomb gaps 
in localized regimes \cite{Efros1975}, Luttinger liquid power laws in 1D \cite{Bockrath1999}, and
Kondo resonances from magnetic moments \cite{Turco2024}—can yield nearly indistinguishable dependences on bias, 
temperature, and magnetic field, obscuring the identification of the underlying mechanism. 
To address this ambiguity, 
we develop a predictive framework that jointly treats strong band anisotropy, weak 
spin-independent disorder, and intrinsic SOC (Rashba/Dresselhaus) in an open-Fermi-surface 
2D conductor. 

Against this backdrop, we focus on the weak-disorder diffusive regime, characterized by 
$k_F l\gg 1$, with mean free path $l=v_F \tau$, where $k_F$ denotes the Fermi wave number 
and $v_F$ the Fermi velocity.
In low-dimensional electronic systems, this regime can qualitatively reshape screening and 
localization. For instance, in 1D and Q1D conductors, it can induce localization phenomena that destabilize the clean-limit DOS. Similarly, in strongly anisotropic systems with Dirac or quasi-Dirac dispersions, weak disorder can generate a finite background DOS through quasi-localized states \cite{Pixley2016}. Such interplay is particularly relevant for open Fermi-surface geometries.

Beyond the generic low-dimensional platforms discussed above, strongly anisotropic 2D 
systems—which we model as a planar array of dislocation tubes with weak, spin-independent 
disorder—are not an abstraction but a realistic substructure in several state-of-the-art 
materials and devices. In III-nitride semiconductors (GaN, InN), parallel threading 
dislocations form arrays of quasi-1D conductive channels that can dominate charge and heat 
transport \cite{Li2020, Yao2025}. Similar dislocation-guided “supermetallic” pathways have 
been identified and are now being engineered inside Si/SiGe MOSFET channels \cite{Reiche2016}. 
In 2D materials, grain boundaries and AB–BA stacking lines in bilayer graphene and hBN realize 
ordered dislocation arrays that host robust 1D metallic modes \cite{Yin2016}. Oxide 
heterostructures (e.g., SrTiO$_3$, LAO/STO bicrystals)
similarly exhibit conduction along 
periodic dislocations and ferroelastic walls, with spacing set by crystal disorientation 
\cite{Thiel2009}. Finally, both theory and experiment on topological insulators and semimetals 
highlight helical 1D modes bound to dislocation lines, underscoring the technological relevance 
of our open–Fermi-surface, anisotropic model \cite{Xue2021}.

To isolate impurity effects in such low-dimensional settings, one works at temperatures low 
enough that thermal fluctuations do not wash out the electron’s accumulated phase during impurity 
scattering, thereby preserving quantum-interference corrections \cite{Bergmann1984, Altland2002}. 
In this regime, electron-electron (e–e) interactions combine with elastic scattering to produce 
interference effects that renormalize single-particle and transport properties. The first 
weak-disorder calculations showed that, already at lowest order in perturbation theory, including 
e e interactions yields a singular correction to the DOS at the Fermi level \cite{Altshuler1979_1, 
Altshuler1980}, with closely related consequences for kinetic and thermodynamic properties 
\cite{Altshuler1985}. Higher-order treatments in the weak-disorder limit established that the form of 
these singularities is preserved while pre-factors are renormalized \cite{Finkelshtein1983, Fukuyama1985}. 
Comparable conclusions were reached for structurally anisotropic 2D systems without SOC, providing a 
natural baseline for the present work where SOC is included explicitly \cite{Firsov1987, Firsov1988}.

Recent experiments and theoretical studies on e-e interactions in weakly disordered conductors consistently show a strong suppression of the DOS near the Fermi energy $\varepsilon_F$. Diagrammatic studies clarify how magnetic fields regularize the low-energy anomaly and modify scaling \cite{Caliskan2012}, while non-equilibrium Green’s function approaches extend e-e interaction corrections to nanoscale devices \cite{Zhou2019}. Scanning tunneling spectroscopy (STS) on strongly disordered ultra-thin films, such as NbN and MoN, has directly observed the Altshuler–Aronov DOS suppression \cite{Carbillet2020, Kuzmiak2023}, and LDOS fluctuation measurements have linked the anomaly to quantum diffusion \cite{Mathieu2023}. On the theory side, topological-insulator surface analyses separate e-e interaction from interference contributions \cite{Shi2023}, and states-conserving AA extensions show how spectral weight depleted near $\varepsilon_F$ is recovered at higher energies \cite{Antonia2018}. More broadly, the interaction–disorder interplay has been connected to Friedel oscillation physics near the 2D metal–insulator transition \cite{Huang2024}, and even in the ballistic regime e-e interaction can reshape the DOS non-perturbatively \cite{Jorg2003}.

In this work, we compute the AA interaction correction to the single-particle DOS for a strongly 
anisotropic 2D metal with an open Fermi surface, weak spin-independent disorder, and intrinsic 
Rashba/Dresselhaus SOC confined to the dispersive $x$ direction. We evaluate the DOS correction 
from the exchange diagrams in the diffusive channel with dynamically screened Coulomb interaction 
within the random-phase approximation
(RPA): the particle–hole bubble is dressed by the impurity ladder, and—after taking the spin trace 
at density vertices—the ladder reduces to the charge/singlet diffuson. 
We neglect the corresponding Hartree contribution, as its requisite large momentum transfer makes it parametrically suppressed for the long-range Coulomb interaction considered here.
The resulting parameter-explicit expression exhibits a clear dimensional crossover governed by the transverse tunneling $t_y$ and the elastic time $\tau$.
The physical role of the SOC strictly depends on this dimensionality. While SOC increases the magnitude of the 2D logarithmic dip, it also remarkably enhances the 1D square-root anomaly. Furthermore, strong SOC in the 1D regime can invert the sign of the correction entirely, producing a positive DOS peak that decays at higher energies.

The rest of the paper is organized as follows. Sec.~\ref{sec:Model} introduces the main model—a strongly 
anisotropic 2D conductor with an open Fermi surface and intrinsic Rashba/Dresselhaus SOC along the 
dispersive $x$ direction—and collects the essential ingredients for the DOS correction: the energy spectrum, 
the disorder model and mean free times $(\tau_1,\tau_2)$, and the explicit forms of the bare Green’s functions 
(in the helicity basis). Sec.~\ref{sec:diag} presents the diagrammatic calculation in the diffusive limit: 
we evaluate the exchange self-energy, compute the particle–hole bubbles, re-sum the impurity ladders to 
obtain the charge (singlet) diffuson $\theta(\mathbf q,i\varepsilon)$, and derive the dynamically screened 
Coulomb interaction $V(\mathbf q,i\varepsilon)$ within RPA. Sec.~\ref{sec:cor} provides analytic expressions 
for the DOS correction $\delta\rho(\varepsilon,T)$ in the 2D and 1D windows, identifies the crossover scale 
$\varepsilon_c\!\sim\! t_y^2\tau$.
Sec.~\ref{sec:Conclusions} summarizes the main results and outlines proposed experimental tests. Finally, For completeness, the explicit derivation of the dynamically screened interaction is detailed in Appendix~\ref{sec:Appendix_A}.


\section{\label{sec:Model}Model and Formalism for the DOS Correction}

\begin{figure}
    \centering
    \includegraphics[width=\linewidth]{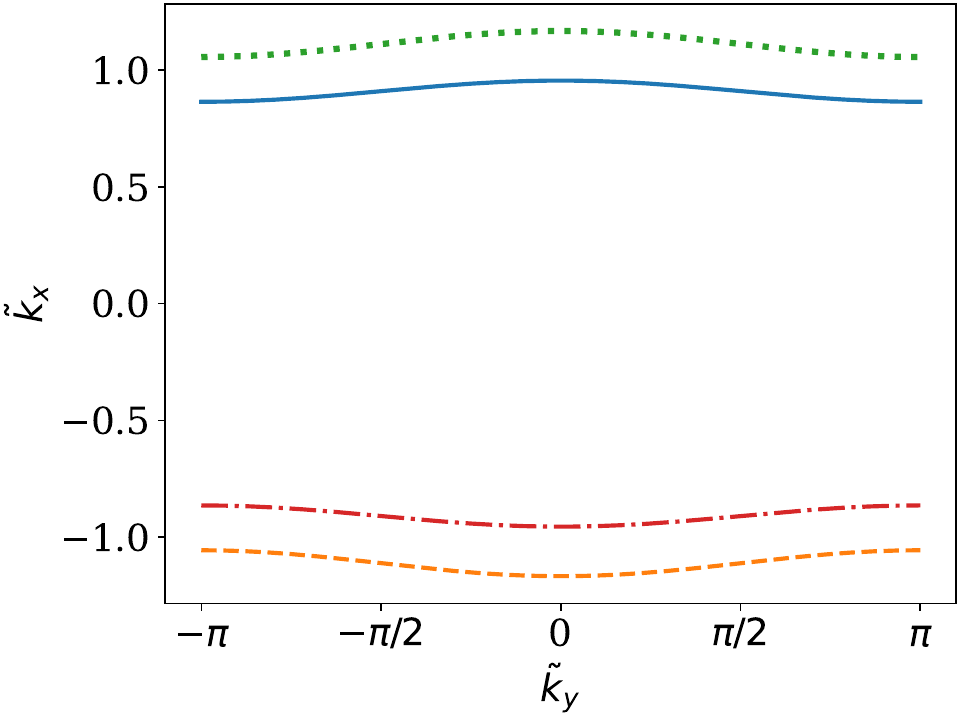}
    \caption{Fermi surface for a strongly anisotropic 2D band. Here $\tilde{k}_x = k_x / k_F$ and $\tilde{k}_y = 
    d k_y$ are dimensionless momenta; parameters used are $2t_y / \left(k_F v_F\right) = 0.05 $ and $\sqrt{\alpha^2 
    + \beta^2}/v_F = 0.1$. }
    \label{fig:Fermi_surface}
\end{figure}

We consider a strongly anisotropic 2D electron system modeled as a planar array of edge-dislocation 
tubes (chains), each tube providing a quasi-1D conducting channel. Electrons propagate predominantly 
along the tubes, while transverse motion occurs only via nearest-neighbor tunneling with amplitude 
$t_y$ between tubes separated by $d$. Intrinsic spin–orbit interaction acts only along the longitudinal 
direction and is taken as $H_{\mathrm{SO}}=\alpha\,k_x\sigma_y-\beta\,k_x\sigma_x$, where $\alpha$ and 
$\beta$ denote the Rashba and Dresselhaus SOC parameters, respectively \cite{Romano2005, 
Suleymanli2023}. The low-energy dispersion near $\varepsilon_F$ is written as
\begin{eqnarray}\label{eq:dis}
    \varepsilon_\sigma(\mathbf{k})&=&\varepsilon_F+v_F\left(\left|k_{x}\right|-k_F\right)
    +\sigma\sqrt{\alpha^2+\beta^2} k_x\nonumber\\
    &&-2t_y \cos \left(
    d k_y\right),
\end{eqnarray}
where $k_x$ is the longitudinal quasi-momentum (along the tubes), $k_y$ is the transverse quasi-momentum, 
and $\sigma=\pm$ labels the helicity. Note that the dispersion in Eq.~(\ref{eq:dis}) represents a spin-orbit-coupled realization of the well-established coupled-wire construction, a mathematical framework widely used to model 2D anisotropic and topological phases \cite{Kane2002, Klinovaja2013}.
We work in the strong-anisotropy regime $v_F>\sqrt{\alpha^2+\beta^2}$ 
and $2t_y/(v_F k_F)\ll1$. Then the helicity-dependent longitudinal velocity $v_x^\sigma=\partial
\varepsilon_\sigma/\partial k_x=v_F\,\mathrm{sgn}(k_x)+\sigma\sqrt{\alpha^2+\beta^2}$ remains positive on 
the right sheet $(k_x>0)$ and negative on the left sheet $(k_x<0)$, preserving the natural chirality of the 
open Fermi lines. The overlap integral $t_y$ decreases exponentially with $d/a$, where $a$ is the tube radius, 
and—within the above inequalities—the Fermi surface consists of four warped lines (two sheets $\times$ two 
helicities), see Fig.~\ref{fig:Fermi_surface}. Throughout we assume the weak-disorder diffusive window 
$k_Fl\gg1$.

In this system, impurities are randomly distributed on the tubes, and spin-independent \cite{Burkov2004}. 
Their potential is effective only within a given tube (no inter-tube scattering). Within the Born approximation, 
scattering decomposes into forward $(q\!\approx\!0)$ and backward $(q\!\approx\!2k_F)$ channels with rates
\begin{equation}
    \tau_1^{-1} = \frac{c_{imp}}{v_F} U^2(0)\quad
    \hbox{and}\quad\tau_2^{-1} = \frac{c_{imp}}{v_F} U^2(2k_F),
    \label{eq:tau}
\end{equation}
respectively, where $c_{\mathrm{imp}}$ is the linear impurity concentration along a tube and $U(q)$ the Fourier 
transform of the impurity potential \cite{Abrikosov2012}. The total momentum-relaxation rate is $\tau^{-1}=
\tau_1^{-1}+\tau_2^{-1}$, and the mean free path is $l=v_F\tau$. 

In this setting, including e-e interactions, the one-particle DOS is obtained from the retarded 
Green's function as
\begin{eqnarray}
    \rho(\varepsilon, T) &=& - \frac{2}{\pi} \int  \frac{d^2 k}{(2\pi)^2} \Im \Bigg( \sum_s \mathrm{Tr}
    \bigg(G_{s}^{-1} \left(\mathbf{k}, i \varepsilon_n \rightarrow \varepsilon + i0^+\right)  \nonumber\\
    && - \Sigma_s\left(\mathbf{k}, i \varepsilon_n \rightarrow \varepsilon + i0^+
    \right) \bigg)^{-1} \Bigg).
    \label{eq:full_dos}
\end{eqnarray}
where the bare Matsubara Green function $G{s}(\mathbf k,i\varepsilon_n)$ characterizes the non-interacting 
problem, $\Sigma_s(\mathbf k,i\varepsilon_n)$ is the self-energy, and $\varepsilon_n=\pi T(2n+1)$ are 
fermionic Matsubara frequencies, $n\in\mathbb Z$. The index $s=\pm$ labels the right $(k_x\!\approx\!+k_F)$ 
and left $(k_x\!\approx\!-k_F)$ open-FS sheets (solid/dotted vs. dashed/dash-dotted lines in 
Fig.~\ref{fig:Fermi_surface}). For our strongly anisotropic spectrum with longitudinal Rashba/Dresselhaus SOC, 
the bare propagator can be written in a $ k$-independent helicity basis as
\begin{eqnarray}
    G_{s} \left(\mathbf{k}, i \varepsilon_n\right) &=& \sum_\sigma \Bigg(i \varepsilon_n -s v_F\left(k_{x}
    -s k_F\right) - \sigma \sqrt{\alpha^2+\beta^2} k_x \nonumber\\
    &&+2t_y \cos \left(d k_y\right) + i/2\tau\ \mathrm{sgn}(\varepsilon_n)\Bigg)^{-1} g_\sigma,
    \label{eq:bare_Green}
\end{eqnarray}
where 
\begin{equation}
    g_\sigma = \frac{1}{2} \begin{pmatrix} 
    1 & \sigma \sqrt{\frac{\beta+i \alpha}{\beta - i \alpha}}\\
    \sigma \sqrt{\frac{\beta - i \alpha}{\beta + i \alpha}} & 1
    \end{pmatrix}.
    \label{eq:g_Green}
\end{equation} 

In this study, we compute the correction to the DOS due to the e-e interaction, where from 
Eq.~(\ref{eq:full_dos}) for this purpose we find for the correction to the DOS 
\begin{eqnarray}
    \delta \rho(\varepsilon, T) &=& -  \frac{2}{\pi} \int \frac{d^2k}{(2\pi)^2}
    \Im \Bigg( \sum_s
    \mathrm{Tr}\bigg[G_s^{2} \left(\mathbf{k}, i \varepsilon_n \rightarrow \varepsilon + i0^+\right)\nonumber\\
    && \times \Sigma_s\left(\mathbf{k}, i \varepsilon_n \rightarrow \varepsilon + i0^+ \right) \bigg]
     \Bigg).
    \label{eq:cor_dos}
\end{eqnarray}
Eq.~(\ref{eq:cor_dos}) represents the leading-order interaction correction. In the weak-disorder regime, higher-order terms are parametrically suppressed and merely renormalize the prefactors without altering the fundamental form of the singularity \cite{Finkelshtein1983, Fukuyama1985}.
This correction can be computed by the conventional diagrammatic technique \cite{Abrikosov2012}, 
and this is presented in detail in Sec.~\ref{sec:diag}.

\section{\label{sec:diag} Impurity vertices and renormalization of the Coulomb interaction}

\begin{figure}
    \centering
    \includegraphics[width=\linewidth]{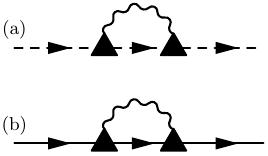}
\caption{Exchange diagrams in the diffusion channel that contribute to $\Sigma_s$ in the 
first order of the e -- e interaction. Dashed (a) and solid (b) arrowed lines represent the bare Green’s functions for the $s=+1$ and $s=-1$ bands, respectively. The wavy line denotes the dynamically screened Coulomb interaction $V_s(\mathbf q,i\omega_m)$, while the filled triangles indicate the impurity ladders $\theta_s(\mathbf q,i\omega_m)$ corresponding to multiple impurity scatterings at the density vertices.}
    \label{fig:diagrams}
\end{figure}

The leading contribution of the e-e interaction self-energy $\Sigma_s(\mathbf{k}, 
i\varepsilon_n)$ to the one-particle DOS arises, to first order in the interaction, from the 
diffusion channel \cite{Finkelshtein1983, Altshuler1985, Fukuyama1985, Firsov1987}. Therefore, 
when evaluating $\Sigma_s(\mathbf{k}, i\varepsilon_n)$ we sum the diagrams shown in 
Fig.~\ref{fig:diagrams}. As already noted in Sec.~\ref{sec:Introduction}, the nontrivial quantum 
corrections driven by e-e interaction occur due to interference between multiple 
impurity scatterings and inelastic e-e processes. Multiple impurity scatterings 
are represented in Fig.~\ref{fig:diagrams} by filled triangles (impurity ladders). In this 
notation, dashed arrowed lines correspond to the bare Green’s function with $s=+1$ (see 
Eq.~(\ref{eq:bare_Green})), while solid lines correspond to the bare Green’s function with $s=-1$.

We compute $\Sigma_s(\mathbf{k}, i\varepsilon_n)$ as
\begin{eqnarray}
    \Sigma_s\left(\mathbf{k}, i \varepsilon_n\right) &=& T  \sum_{\omega_{\mathrm{m}}} 
    \int \frac{\mathrm{~d}^2 q}{(2 \pi)^2} G_s\left(\boldsymbol{k}+\boldsymbol{q}, i\left(\varepsilon_n+
    \omega_{\mathrm{m}}\right)\right) \nonumber\\
    && \times \theta_s^2\left(\boldsymbol{q}, i \omega_{\mathrm{m}}\right) 
    V_{s}\left(\boldsymbol{q}, i \omega_{\mathrm{m}}\right),
    \label{eq:Sigma}
\end{eqnarray}
where $T$ is the temperature, $\omega_m=2\pi T m$ is the transferred Matsubara frequency, 
and $\mathbf q$ is the transferred momentum in the inelastic e-e scattering. 
The factor $\theta_s(\mathbf q,i\omega_m)$ denotes the impurity ladder 
(i.e., the dressed density vertex) represented by triangles in Fig.~\ref{fig:diagrams}, 
while $V_s(\mathbf q,i\omega_m)$ is the screened Coulomb interaction, depicted by the 
wavy line in Fig.~\ref{fig:diagrams}. 

\begin{figure}
\centering
\includegraphics[width=\linewidth]{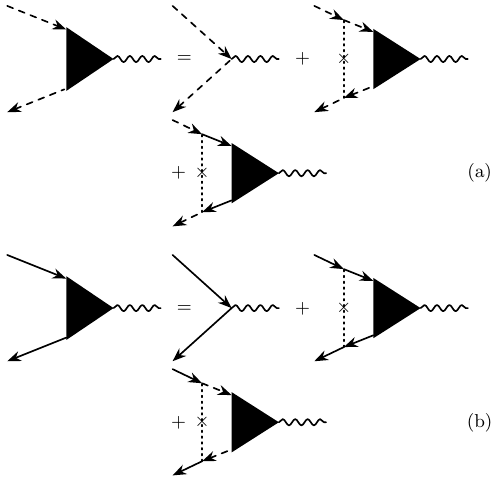}
\caption{Ladder approximation for calculating impurity vertices.}
\label{fig:vertex}
\end{figure}

Here we outline the calculation of the impurity ladders. We work in the ladder approximation 
because we are in the weak-disorder regime, where diagrams with intersecting impurity lines 
are suppressed by the small parameter $(k_Fl)^{-1}\!\ll\!1$ \cite{Abrikosov2012}. The 
renormalization of the vertices by impurity lines is implemented via the graphical Dyson 
equations shown in Fig.~\ref{fig:vertex}: Fig.~\ref{fig:vertex}a corresponds to the 
impurity ladder relevant for the diagram in Fig.~\ref{fig:diagrams}a, and 
Fig.~\ref{fig:vertex}b to the one for Fig.~\ref{fig:diagrams}b. From Fig.~\ref{fig:vertex} 
we can write

\begin{eqnarray}
    \theta_1\left(\boldsymbol{q}, i \omega_{\mathrm{m}}\right) &=& 1 
    + \frac{d v_F}{\tau_1} I_1 (\mathbf{q}, i\omega_m) \theta_1\left(\boldsymbol{q}, i 
    \omega_{\mathrm{m}}\right)\nonumber\\ 
    &&+ \frac{d v_F}{\tau_2} I_2 (\mathbf{q}, i\omega_m) \theta_2\left(\boldsymbol{q}, 
    i \omega_{\mathrm{m}}\right), 
    \label{eq:alpha_1}
\end{eqnarray}

\begin{eqnarray}
    \theta_2\left(\boldsymbol{q}, i \omega_{\mathrm{m}}\right) &=& 1 
    + \frac{d v_F}{\tau_1} I_2 (\mathbf{q}, i\omega_m) \theta_2\left(\boldsymbol{q}, 
    i \omega_{\mathrm{m}}\right)\nonumber\\ 
    &&+ \frac{d v_F}{\tau_2} I_1 (\mathbf{q}, i\omega_m) \theta_1\left(\boldsymbol{q}, 
    i \omega_{\mathrm{m}}\right), 
    \label{eq:alpha_2}
\end{eqnarray}
where $\tau_{1,2}$ are defined in Eq.~(\ref{eq:tau}), and $I_{1,2}$ are the elementary 
bubbles from which the ladders are built (displayed in Fig.~\ref{fig:bubbles}a and b, 
respectively). We begin by evaluating $I_1$ (Fig.~\ref{fig:bubbles}a):

\begin{figure}
\centering
\includegraphics[width=\linewidth]{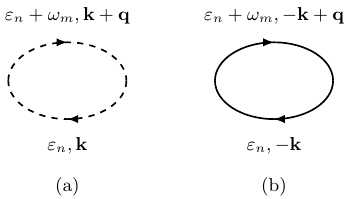}
\caption{Elementary bubbles.}
\label{fig:bubbles}
\end{figure}

\begin{eqnarray}
    I_1(\mathbf{q}, i\omega_m) &=& \int \frac{d^2k}{(2\pi)^2} \mathrm{Tr}\bigg[ 
    G_1\left(\boldsymbol{k}+\boldsymbol{q}, i\left(\varepsilon_n+
    \omega_{\mathrm{m}}\right)\right)\nonumber\\
    && \times G_1\left(\boldsymbol{k}, i \varepsilon_n\right)\bigg],
    \label{eq:I_1_main}
\end{eqnarray}
where $G_1\left(\boldsymbol{k}, i \varepsilon_n\right)$ taken from 
Eq.~(\ref{eq:bare_Green}) for $s=+1$. From Eqs.~(\ref{eq:g_Green}) and (\ref{eq:I_1_main}) 
we find 
\begin{equation}
    \mathrm{Tr} \left[g_\sigma g_{\sigma^\prime}\right] = \delta_{\sigma\sigma^\prime}.
\end{equation}
Moreover, using the residue theorem one sees that $I_1=0$ whenever $\varepsilon_n
(\varepsilon_n+\omega_m)>0$ \cite{Arfken2013}. Therefore, in what follows we restrict to 
$\varepsilon_n(\varepsilon_n+\omega_m)<0$ and treat the two cases separately — (i) 
$\varepsilon_n<0$ with $\varepsilon_n+\omega_m>0$ and (ii) $\varepsilon_n>0$ with 
$\varepsilon_n+\omega_m<0.$ Evaluating each case explicitly yields
\begin{eqnarray}
    I_1(\mathbf{q}, i\omega_m) &=& \sum_\sigma \frac{\tau}{d v_F^\sigma} 
    \Bigg(  1 - |\omega_m| \tau - i l^\sigma q_x \mathrm{sgn} (\omega_m)  \nonumber\\
    && - (l^\sigma q_x)^2 - 4\tau^2 t_y^2 \left(1 - \cos(d q_y)\right)\Bigg), 
    \label{eq:I_1}
\end{eqnarray} 
where $v_F^{\sigma}=v_F+\sigma\sqrt{\alpha^2+\beta^2}$ and $l^{\sigma}=v_F^{\sigma}\tau$. 
In deriving Eq.~(\ref{eq:I_1}) we use the diffusive small-parameter conditions 
$|\omega_m|\tau\ll1, q_x l^{\sigma}\ll1,$ and $t_y\tau\ll1$. Physically, these inequalities 
mean that within the inelastic time scale $\sim 1/|\omega_m|$, and before a transverse 
tunneling event, an electron undergoes many elastic impurity scatterings, while the 
longitudinal momentum transfer $q_x$ remains small enough that the motion stays diffusive.
Note that these inequalities also establish the validity limits for all calculations in this study.

For $I_2(\mathbf q,i\omega_m)$ we proceed analogously, now using the bare Green’s function 
of Eq.~(\ref{eq:bare_Green}) with $s=-1$, which yields
\begin{equation}
    I_2(\mathbf{q}, i\omega_m) = I_1^*(\mathbf{q}, i\omega_m).
    \label{eq:I_2}
\end{equation}
Combining Eq.~(\ref{eq:I_2}) with Eqs.~(\ref{eq:alpha_1}) and (\ref{eq:alpha_2}) we obtain
\begin{eqnarray}
    \theta_1\left(\boldsymbol{q}, i \omega_{\mathrm{m}}\right) &=& \bigg(1
    +(\tau_2^{-1} - \tau_1^{-1})d v_F I_1^*\bigg) \nonumber\\
    &&\times \bigg(1-2\tau_1^{-1}d v_F \Re(I_1) \nonumber\\
    &&- \tau^{-1} (\tau_2^{-1} - \tau_1^{-1}) d^2 v_F^2 I_1I_1^*\bigg)^{-1}. \nonumber\\
    \label{eq:theta_ara}
\end{eqnarray}
Finally, substituting Eq.~(\ref{eq:I_1}) into Eq.~(\ref{eq:theta_ara}) gives
\begin{equation}
    \theta_1\left(\boldsymbol{q}, i \omega_{\mathrm{m}}\right) = \sum_
    \sigma \frac{\theta_{11}\left(\boldsymbol{q}, i \omega_{\mathrm{m}}\right)}
    { \theta_{12}\left(\boldsymbol{q}, i \omega_{\mathrm{m}}\right)},
    \label{eq:alpha_main_1}
\end{equation}
where
\begin{eqnarray}
    &&\theta_{11}\left(\boldsymbol{q}, i \omega_{\mathrm{m}}\right) = 1 + 
    \tau\left(\tau_2^{-1}-\tau_1^{-1}\right) \frac{v_F}{v_F^\sigma}\nonumber\\
    && \times \Bigg(  1 - |\omega_m| \tau + i l^\sigma q_x \mathrm{sgn} (\omega_m)  \nonumber\\
    && - (l^\sigma q_x)^2 - 4\tau^2 t_y^2 \left(1 - \cos(d q_y)\right)\Bigg) ,
\end{eqnarray}
\begin{eqnarray}
    &&\theta_{12}\left(\boldsymbol{q}, i \omega_{\mathrm{m}}\right) =
    1-2\tau\tau_1^{-1}\frac{v_F}{v_F^\sigma}\Bigg(  1 - |\omega_m| \tau  \nonumber\\
    && - 4\tau^2 t_y^2 \left(1 - \cos(d q_y)\right)\Bigg)+\tau(\tau_2^{-1} - \tau_1^{-1}) 
    \left(\frac{v_F}{v_F^\sigma}\right)^2 \nonumber\\
    && \times \Bigg(1-2|\omega_m|\tau - (l^\sigma q_x)^2 - 8 \tau^2 t_y^2 \left(1 - 
    \cos(d q_y)\right)\Bigg) \nonumber\\
\end{eqnarray}
and 
\begin{equation}
    \theta_2\left(\boldsymbol{q}, i \omega_{\mathrm{m}}\right) = 
    \theta_1^*\left(\boldsymbol{q}, i \omega_{\mathrm{m}}\right). 
    \label{eq:alpha_main_2}
\end{equation}

We now compute the screened Coulomb interaction $V_{s}(\mathbf q,i\omega_m)$. Since the bare 
Coulomb potential is long-ranged, its renormalization by electronic polarization must be 
taken into account \cite{Hamaguchi2009}. Employing the RPA is necessary here to capture this collective screening and prevent unphysical divergences as $q \rightarrow 0$.
The Dyson equation for the dynamically screened interaction can be written as
\begin{equation}\label{eq:Vs_first}
    V_{s}\left(\boldsymbol{q}, i \omega_{\mathrm{m}}\right) = V_0 (\mathbf{q}) + 
    V_0 (\mathbf{q}) \Pi_s \left(\boldsymbol{q}, i \omega_{\mathrm{m}}\right) 
    V_{s}\left(\boldsymbol{q}, i \omega_{\mathrm{m}}\right),
\end{equation}
where the bare 2D Coulomb kernel is $V_0(\mathbf q)=2\pi e^2/\sqrt{q_x^2+q_y^2}$ 
\cite{Altshuler1985, Firsov1987}, and $\Pi_s(\mathbf q,i\omega_m)$ is the polarization 
operator, evaluated as
\begin{eqnarray}
    \Pi_s \left(\boldsymbol{q}, i \omega_{\mathrm{m}}\right) &=& T \int \frac{d^2k}{(2\pi)^2} 
    \sum_{\varepsilon_n} \mathrm{Tr}[G_s\left(\boldsymbol{k}+\boldsymbol{q}, i\left(\varepsilon_n+
    \omega_{\mathrm{m}}\right)\right) \nonumber\\
    && \times G_s\left(\boldsymbol{k}, i \varepsilon_n\right)]\ \theta_s\left(\boldsymbol{q}, 
    i \omega_{\mathrm{m}}\right).
    \label{eq:polarizasyon}
\end{eqnarray}
Note that when computing $\Pi_s(\mathbf q,i\omega_m)$, we perform the Matsubara sum over 
$\varepsilon_n$ before the $\mathbf k$-integration. If one integrates over $\mathbf k$ first, 
the result would be nonzero only for $\varepsilon_n(\varepsilon_n+\omega_m)<0$, incorrectly 
discarding other frequency sectors. 
Consequently, Matsubara sum naturally splits into three parts,
\begin{equation}
    \sum_{\varepsilon_n} =  
    \sum_{\varepsilon_n = - \omega_m}^0 + \sum_{\varepsilon_n = 0}^\infty + \sum_{\varepsilon_n = - \omega_m}^{-\infty},
\end{equation}
and only the first sum $\sum_{\varepsilon_n=-\omega_m}^{0}$ acquires impurity-ladder 
corrections through $\theta_s(\mathbf q,i\omega_m)$; for the other two sums 
one has $\theta_s=1$ because the corresponding bubbles $I_{1,2}$ vanish. Evaluating each sum using
\begin{equation}\label{eq: help_int}
    T \sum_{\varepsilon_n} \rightarrow \frac{1}{2 \pi i} \int_C d \varepsilon \frac{\mathrm{sgn}\, 
    \varepsilon}{\exp\left(\left(\varepsilon - \varepsilon_F\right)/T\right) + 1}
\end{equation}
and then performing the $\mathbf k$-integration via the residue theorem \cite{Altland2010} (see Appendix~\ref{sec:Appendix_A} for explicit details of the integration contours and derivations), 
we obtain
\begin{equation}
    V_{s}\left(\boldsymbol{q}, i \omega_{\mathrm{m}}\right) = 2 \pi e^2 \left(|\mathbf{q}| 
    + 2\pi e^2 \sum_\sigma \rho_0^\sigma \frac{V_1^\sigma\left(\boldsymbol{q}\right)}{|\omega_m| 
    + V_1^\sigma\left(\boldsymbol{q}\right)} \right)^{-1},
    \label{eq:V_main}
\end{equation}
where 
\begin{eqnarray}
    V_1^\sigma\left(\boldsymbol{q}\right) &=& \frac{\left(v_F^\sigma\right)^2 \tau_2}{2} 
    q_x^2 + 4 \tau t_y^2 (1-\cos(d q_y)),
    \label{eq:V1}
\end{eqnarray}
and $ \rho_0^\sigma = 2/\left(\pi v_F^\sigma d\right)$.

\section{\label{sec:cor} Correction to the DOS}

Substituting Eqs.~(\ref{eq:Sigma}), (\ref{eq:alpha_main_1}), (\ref{eq:alpha_main_2}), and (\ref{eq:V_main}) 
into Eq.~(\ref{eq:cor_dos}) yields the exchange-diagram correction of Fig.~\ref{fig:diagrams} to the DOS. 
For the Matsubara sum $\sum_{\omega_m}$ appearing in Eq.~(\ref{eq:Sigma}), we convert the $\omega_m$ sum to 
a contour integral and then to an integral along the imaginary axis; at this point we set $i\varepsilon_n
\to \varepsilon$ (analytic continuation). Because $\varepsilon_n=\pi T(2n+1)$ and $\omega_m=2\pi T m$, the 
integration contour does not pass through the origin; therefore we introduce $\varepsilon_l=\omega_m+
\varepsilon_n$ (with $l=n+m$). When replacing the sum by an integral, we use the standard identity for 
fermionic sums, $\mathrm{Res}\,[\tanh(\varepsilon/2T)]=2T$ \cite{Altland2010}. As a result, from Eq.~
(\ref{eq:cor_dos}) we obtain
\begin{eqnarray}
    \delta \rho (\varepsilon, T) &=& - 2 e^2 \sum_\sigma \rho_0^\sigma \int d \omega  \frac{ \sinh 
    \frac{\varepsilon+\omega}{T}}{\cosh \frac{\varepsilon+\omega}{T} - \cosh \frac{\varepsilon-\omega}{T}}
    \nonumber\\
    && \times \Im \Bigg(\int \frac{d^2q}{(2\pi)^2} \left(-i\omega + V_1^\sigma\left(\boldsymbol{q}\right) 
    \right)^{-1} \nonumber\\
    && \times \left(-i\omega |\mathbf{q}| + 2 \pi e^2 \rho_0^\sigma V_1^\sigma\left(\boldsymbol{q}\right) 
    \right)^{-1} \Bigg).
    \label{eq:cor_final_1}
\end{eqnarray}
Substituting Eq.~(\ref{eq:V1}) in Eq.~(\ref{eq:cor_final_1}) explicitly, changing variables, and specifying 
the integration domains, we find
\begin{equation}
    \delta \rho (\varepsilon, T) = - \sum_\sigma \frac{\tau}{\pi^3 d l_2^\sigma} \int_0^1 \frac{d \xi}{\xi^2}  
    \frac{ \Im(A(\xi))\sinh \frac{\varepsilon+\xi/\tau}{T}}{\cosh \frac{\varepsilon+\xi/\tau}{T} - 
    \cosh \frac{\varepsilon-\xi/\tau}{T}},
    \label{eq:cor_final_2}
\end{equation}
where
\begin{eqnarray}
    A(\xi) &=& \int_0^1 dx \int_0^1 dy \left( -i +  c_1 x^2 + c_2 y^2\right)^{-1} \nonumber\\
    && \times \left( -i \sqrt{c_3^2 x^2 + c_4^2 y^2} + c_1 x^2 + c_2 y^2\right)^{-1}.
    \label{eq:A(xi)}
\end{eqnarray}
and $\xi = \omega \tau, x = l_2^\sigma q_x, y = dq_y/2, c_1 = 2\tau \tau_2^{-1} \xi^{-1}, c_2 = 
8 t_y^2 \tau^2 \xi^{-1}, c_3 = \left(2 \pi e^2 \rho_0^\sigma l_2\right)^{-1}$ and $c_4 = 
\left(\pi e^2 \rho_0^\sigma d\right)^{-1}.$
As is evident, to compute the DOS correction we first perform the $\mathbf q$-integration—this 
defines $A(\xi)$ as in Eq.~(\ref{eq:A(xi)})—and then integrate over $\omega$; all steps are carried 
out below. Our calculations show that the $\xi$-dependence of $A(\xi)$ changes with $c_2$, so we 
analyze the two asymptotic regimes separately, $c_2\gg 1$ and $c_2\ll 1$, in the subsections 
that follow.

\begin{figure}
    \centering
    \includegraphics[width=0.48\textwidth]{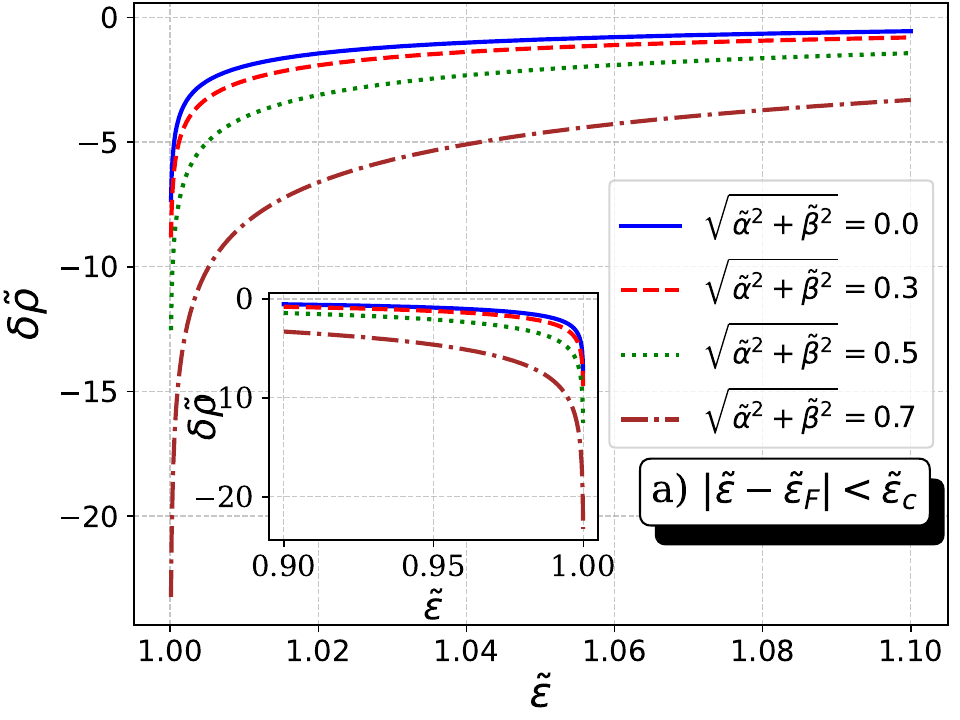}
    \hfill 
    \includegraphics[width=0.48\textwidth]{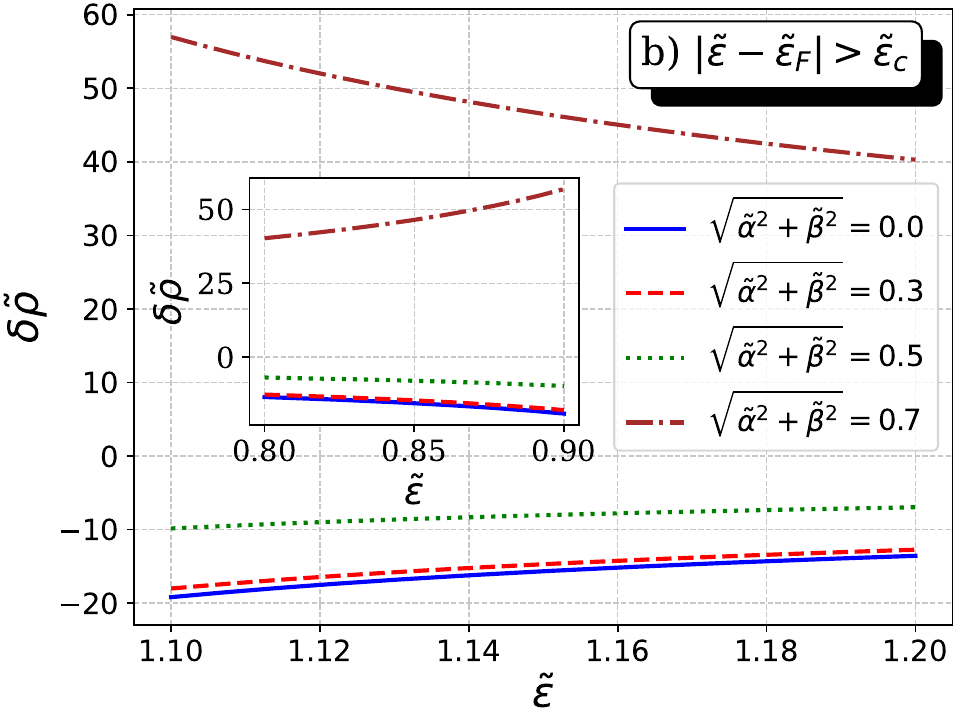}
    \caption{Effects of Rashba and Dresselhaus SOCs on the interaction-induced correction to the density of states in strongly anisotropic 2D systems. (a) The $|\varepsilon - \varepsilon_F| < \varepsilon_c$ regime, plotted using Eq.~(\ref{eq:final_1}). (b) The $|\varepsilon - \varepsilon_F| > \varepsilon_c$ regime, plotted using Eq.~(\ref{eq:final_2}). Here, the following dimensionless quantities are used: $\tilde{\varepsilon} = \varepsilon\tau$, $\tilde{\varepsilon}_{F,c} = \varepsilon_{F,c}\tau$, $\delta\tilde{\rho}(\tilde{\varepsilon}) = \delta{\rho}(\tilde{\varepsilon}) v_F d$, $\tilde{\alpha} = \alpha / v_F$, and $\tilde{\beta} = \beta / v_F$. The Fermi energy and the critical energy are chosen as $\varepsilon_F = 1/\tau$ and $\varepsilon_c = 1/(10\tau)$. All curves are plotted in the strictly zero-temperature limit ($T=0$) to isolate the pure energy dependence of the anomalies.}
    \label{fig:dos}
\end{figure}

\underline{1) $c_2 \gg 1$ (i.e., $\omega \ll 8 t_y^2 \tau$).} We pass to polar coordinates, 
$x=r\cos\varphi$ and $y=r\sin\varphi$, so that Eq.~(\ref{eq:A(xi)}) takes the form
\begin{equation}
    A(\xi) = \frac{1}{2} \int_0^\pi d \varphi \Phi^4_1(\varphi) \int_0^1 dr (r - i 
    \Phi_2(\varphi))^{-1} (r^2 - i \Phi_1^2(\varphi))^{-1}
    \label{eq:A11}
\end{equation}
where 
\begin{equation}
    \Phi_1(\varphi) = (c_1\cos^2\varphi + c_2 \sin^2\varphi)^{-1/2}, 
    \label{eq:Phi_1}
\end{equation}
\begin{equation}
    \Phi_2(\varphi) = \Phi_1^{2}(\varphi) \sqrt{c_3^2 \cos^2 \varphi + c_4^2 \sin^2 \varphi}.
    \label{eq:Phi_2}
\end{equation}
From Eq.~(\ref{eq:A11}) we obtain $A(\xi)$ by integrating Eq.~(\ref{eq:A11}) first over $r$ and 
then over $\varphi$. Integrating Eq.~(\ref{eq:A11}) with respect to $r$ yields
\begin{eqnarray}
    && \int_0^1 dr (r - i \Phi_2)^{-1} (r^2 - i \Phi_1^2)^{-1} \nonumber\\
    && = \frac{1}{\Phi_1^4+\Phi_2^4} \Bigg[ \frac{\Phi_2^2-i \Phi_1^2}{4} \ln \left(\frac{1+
    \Phi_1^{-4}}{\left( 1+\Phi_2^{-2}\right)^2}\right) \nonumber\\
    && + \frac{i\Phi_2^2 + \Phi_1^2}{2} \left(\arctan\left(\frac{1}{\Phi_1^2}\right) - 
    2\arctan\left(\frac{1}{\Phi_2}\right) \right)\nonumber\\
    &&+ 2^{-5/2} \frac{\Phi_2 (\Phi_1^2+\Phi_2^2) + i \Phi_2 (\Phi_2^2 - \Phi_1^2)}{\Phi_1}\nonumber\\
    &&\times \ln\left(\frac{1-\sqrt{2}\Phi_1 +\Phi_1^2}{1+\sqrt{2}\Phi_1 +\Phi_1^2}\right)\nonumber\\
    &&- 2^{-3/2} \frac{\Phi_2 (\Phi_1^2-\Phi_2^2) - i \Phi_2 (\Phi_2^2 + \Phi_1^2)}{\Phi_1}\nonumber\\
    && \times \bigg(\arctan \left(\frac{1}{\sqrt{2}\Phi_1 - 1}\right)\nonumber\\
    &&+ \arctan \left(\frac{1}{\sqrt{2}\Phi_1 + 1}\right) \bigg)
    \Bigg].
    \label{eq:Ar}
\end{eqnarray}
Substituting Eq.~(\ref{eq:Ar}) back into Eq.~(\ref{eq:A11}) allows us to perform the $\varphi$-integration. 
In the regime $c_2\gg 1$, Eq.~(\ref{eq:Ar}) simplifies substantially; inserting this simplification into 
Eq.~(\ref{eq:A11}) we arrive at an expression for $\Im \left(A(\xi)\right)$:
\begin{equation}
    \Im \left(A(\xi)\right) = - \frac{1}{2} \int_0^\pi d\varphi \Phi_1^2(\varphi) \ln \left(\frac{
    \Phi_2(\varphi)}{\Phi_1(\varphi)}\right).
    \label{eq:Ac1>>1}
\end{equation}
Using Eqs.~(\ref{eq:Phi_1}) and (\ref{eq:Phi_2}) in Eq.~(\ref{eq:Ac1>>1}), evaluation of the integral gives
\begin{equation}
    \Im \left(A(\xi)\right) = - \frac{1}{2} \frac{\pi}{\sqrt{c_1 c_2}} \ln\left(\frac{1}{2}\left(
    \frac{c_3}{\sqrt{c_1}} + \frac{c_4}{\sqrt{c_2}} \right)\right).
    \label{eq:A_asimp_final_1}
\end{equation}
Finally, substituting Eq.~(\ref{eq:A_asimp_final_1}) into Eq.~(\ref{eq:cor_final_2}) and integrating 
over $\xi$, we obtain
\begin{eqnarray}
    \delta \rho (\varepsilon, T) &=& - \frac{\ln\Bigl(|\tau \bigl(\max\{ \varepsilon, T\}-\varepsilon_F\bigl)
    |\Bigl)}{8\pi^2 d \sqrt{\tau \tau_2}t_y\left(v_F^2 - \left(\alpha^2+\beta^2\right)\right)} \nonumber\\
    && \times \bigg(v_F \left[\ln\Bigl(|\tau \bigl(\max\{ \varepsilon, T\}-\varepsilon_F\bigl)|\Bigl) 
    + 4  \Lambda_1\right] \nonumber\\
    && + \sqrt{\alpha^2+\beta^2} \Lambda_2\bigg),
    \label{eq:final_1}
\end{eqnarray}
where
\begin{equation}
    \Lambda_1 = \ln\frac{v_F^2 - \left(\alpha^2+\beta^2\right) + 4 d \sqrt{\tau \tau_2^{-1}} t_y 
    \left(v_F + d \sqrt{\tau \tau_2^{-1}} t_y\right)}{128 e^4 \tau^2 t_y^2},
\end{equation}
\begin{eqnarray}
    &&\Lambda_2 = 4 \ln \frac{v_F - \sqrt{\alpha^2+\beta^2}+d \sqrt{\tau\tau_2^{-1}}t_y}{v_F + 
    \sqrt{\alpha^2+\beta^2} +d \sqrt{\tau\tau_2^{-1}}t_y}.
\end{eqnarray}
Eq.~(\ref{eq:final_1}) shows that there is a critical energy $\varepsilon_c = 8 t_y^2 \tau$ such that 
for $|\varepsilon-\varepsilon_F|\!<\!\varepsilon_c$ the DOS exhibits the two-dimensional, logarithmic 
singularity at the Fermi level. Spin–orbit coupling does not shift this critical scale but suppresses 
the overall amplitude of the anomaly. This behavior is illustrated in Fig.~\ref{fig:dos}a for the energy range $(\tilde{\varepsilon}_F,\ \tilde{\varepsilon}_F+\tilde{\varepsilon}_c)$, while the corresponding behavior for the range $(\tilde{\varepsilon}_F-\tilde{\varepsilon}_c,\ \tilde{\varepsilon}_F)$ is shown in the inset.
Here, $\tilde{\varepsilon}_{F,c} = \varepsilon_{F,c} \tau$.

\underline{2) $c_2 \ll 1$ (i.e., $\omega \gg 8 t_y^2 \tau$).} To treat this regime, we first 
extract $\Im (A(\xi))$ from Eq.~(\ref{eq:A(xi)}):
\begin{eqnarray}
    &&\Im (A(\xi)) = \int_0^1 dx \int_0^1 dy\nonumber\\
    && \frac{\left(c_1 x^2 + c_2 y^2\right) \left(1+\sqrt{c_3^2 x^2 + c_4^2 y^2}\right)}
    {\left(\left(c_1 x^2 +c_2 y^2\right)^2+c_3^2 x^2 +c_4^2 y^2\right)\left(1+\sqrt{c_1^2 x^2 
    + c_2^2 y^2}\right)}\nonumber\\
    \label{eq:ImAxi}
\end{eqnarray}
Specializing Eq.~(\ref{eq:ImAxi}) to $c_2\ll 1$, we integrate over $x$ on the 
interval $(0,\,1/\sqrt{c_1})$ and then perform the $y$-integration, which yields
\begin{equation}
    \Im (A(\xi)) \sim \frac{1}{\sqrt{c_1} c_4 \xi^2} \left(\frac{2}{9} + \frac{\pi}{2} 
    + \frac{\ln c_4}{3} \right),
\end{equation}
On the complementary interval $(1/\sqrt{c_1},\,1)$, the integration yields
\begin{equation}
    \Im (A(\xi)) \sim \frac{1}{\sqrt{c_1} c_4 \xi^2} \left(\frac{\pi}{6} - 2 + \ln c_4 \right).
\end{equation}
Hence, we can write
\begin{equation}
    \Im (A(\xi)) = \frac{1}{3\sqrt{c_1} c_4 \xi^2} \left(2\pi - \frac{16}{3} + 4 \ln c_4\right) 
    + O(c_2),
    \label{eq:Ac4>1}
\end{equation}
where $O(c_2)$ denotes terms that are proportional to $c_2$ and thus parametrically small in 
this limit. 
Finally, substituting Eq.~(\ref{eq:Ac4>1}) into Eq.~(\ref{eq:cor_final_2}) 
and carrying out the remaining integration, we obtain
\begin{equation}
    \delta \rho (\varepsilon) = - \frac{4e^2\Lambda_3}{\left(v_F^2 - 
    (\alpha^2+\beta^2)\right)^2|\tau \left(\varepsilon-\varepsilon_F\right)|^{1/2}},
    \label{eq:final_2}
\end{equation}
where
\begin{eqnarray}
    \Lambda_3 &=&  \left(v_F^2 + (\alpha^2 + \beta^2) \right) \left(2 + \ln\left( 
    \frac{\pi^2 (v_F^2 - (\alpha^2 + \beta^2))}{16 e^4}\right)\right) \nonumber\\
    &&+ 2 v_F \sqrt{\alpha^2+\beta^2} \ln \left(\frac{v_F - \sqrt{\alpha^2+\beta^2}}
    {v_F + \sqrt{\alpha^2+\beta^2}}\right).
\end{eqnarray}
As seen from Eq.~(\ref{eq:final_2}), in the regime $|\varepsilon-\varepsilon_F|>\varepsilon_c$ the 
interaction-induced DOS correction crosses over to the stronger, one-dimensional square-root–type 
singularity. 
In Fig.~\ref{fig:dos}b, this correction is plotted for energies greater than $\tilde{\varepsilon}_F+\tilde{\varepsilon}_c$, while the inset displays energies smaller than $\tilde{\varepsilon}_F-\tilde{\varepsilon}_c$. 
As evident from Fig.~\ref{fig:dos}b, the dimensionless interaction correction, defined from Eq.~(\ref{eq:final_2}) as $\delta \tilde{\rho} = v_F d \delta \rho$, attains larger values with increasing Rashba and Dresselhaus SOIs.
This indicates that for $|\varepsilon-\varepsilon_F|>\varepsilon_c$, the DOS correction not only crosses over from a 2D logarithmic to a 1D square-root singularity, but the underlying mechanism of the SOIs also changes. Specifically, the SOIs, which suppress the DOS correction in the $|\varepsilon-\varepsilon_F|<\varepsilon_c$ regime, cause it to take larger values when $|\varepsilon-\varepsilon_F|>\varepsilon_c$. Furthermore, Fig.~\ref{fig:dos}b reveals that under the influence of the SOIs, $\delta\tilde{\rho}$ can change its sign and assume positive values. The dependence of $\delta\tilde{\rho}$ on $\tilde{\varepsilon}$ also changes character at these positive values. Namely, for energies greater than $\tilde{\varepsilon}_F+\tilde{\varepsilon}_c$, $\delta\tilde{\rho}$ increases at higher energies when $\delta \tilde{\rho} < 0$. However, when $\delta\tilde{\rho}$ is positive, it decreases with increasing energy. A similar behavior is valid for the energies smaller than $\tilde{\varepsilon}_F-\tilde{\varepsilon}_c$ shown in the inset of Fig.~\ref{fig:dos}b. It is also useful to emphasize that, since Eqs.~\eqref{eq:final_1} and \eqref{eq:final_2} are asymptotic expressions obtained from the same integral representation in the opposite limits $c_2\gg 1$ and $c_2\ll 1$, respectively, they describe the DOS correction sufficiently below and sufficiently above the crossover scale, but do not constitute a uniform description of the crossover region. Consequently, exact matching at
$\tilde{\varepsilon}=\tilde{\varepsilon}_F\pm\tilde{\varepsilon}_c$
is not imposed in Figs.~\ref{fig:dos}a and \ref{fig:dos}b.

\section{Renormalized DOS and Perturbative Validity}

\begin{figure}
    \centering
    \includegraphics[width=0.48\textwidth]{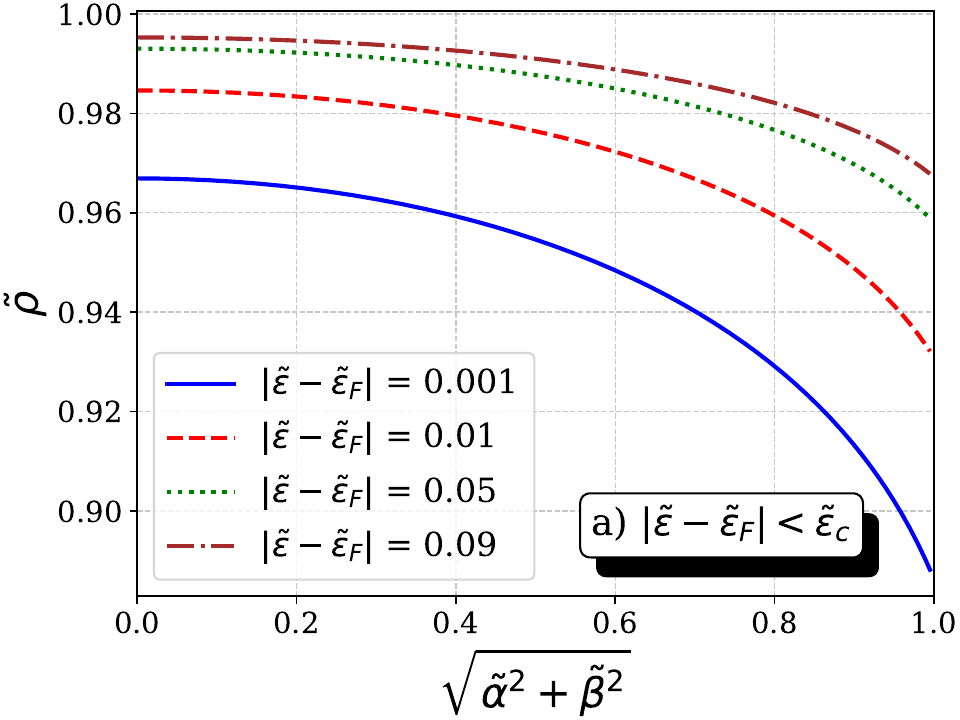}
    \hfill 
    \includegraphics[width=0.48\textwidth]{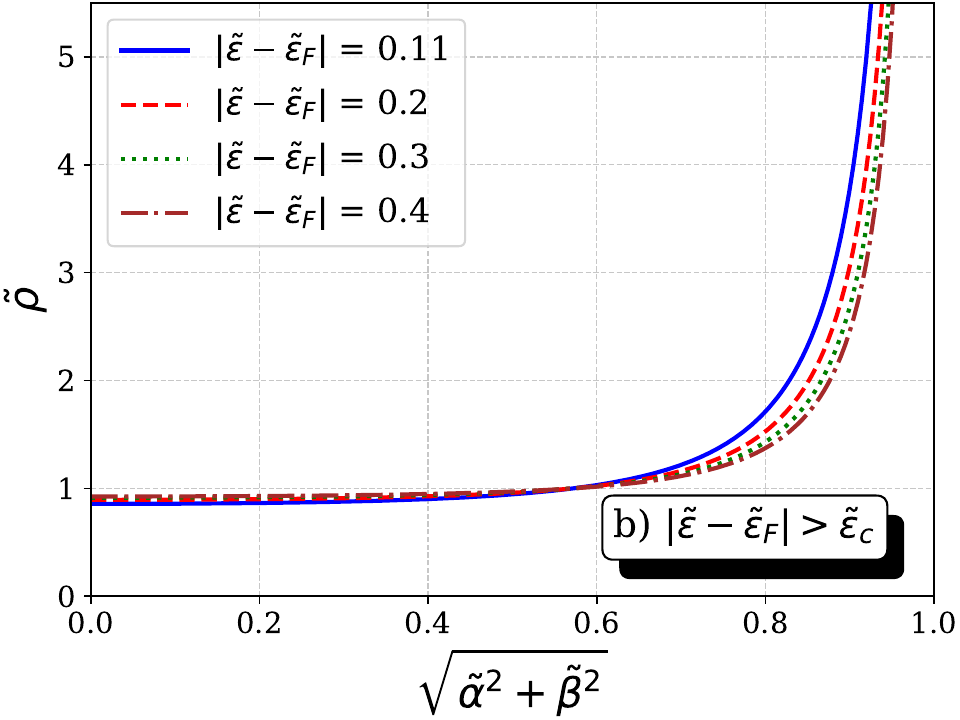}
    \caption{Evolution of the dimensionless density of states per transverse conducting channel $\tilde{\rho}$ with effective spin-orbit coupling strength $\sqrt{\tilde{\alpha}^2 + \tilde{\beta}^2}$ for various energy distances $|\tilde{\varepsilon}-\tilde{\varepsilon}_F|$ from the Fermi level. Here, the critical energy is chosen as $\varepsilon_c = 1/(10\tau$).}
    \label{fig:full_DOS}
\end{figure}

\begin{figure}
    \centering
    \includegraphics[width=0.48\textwidth]{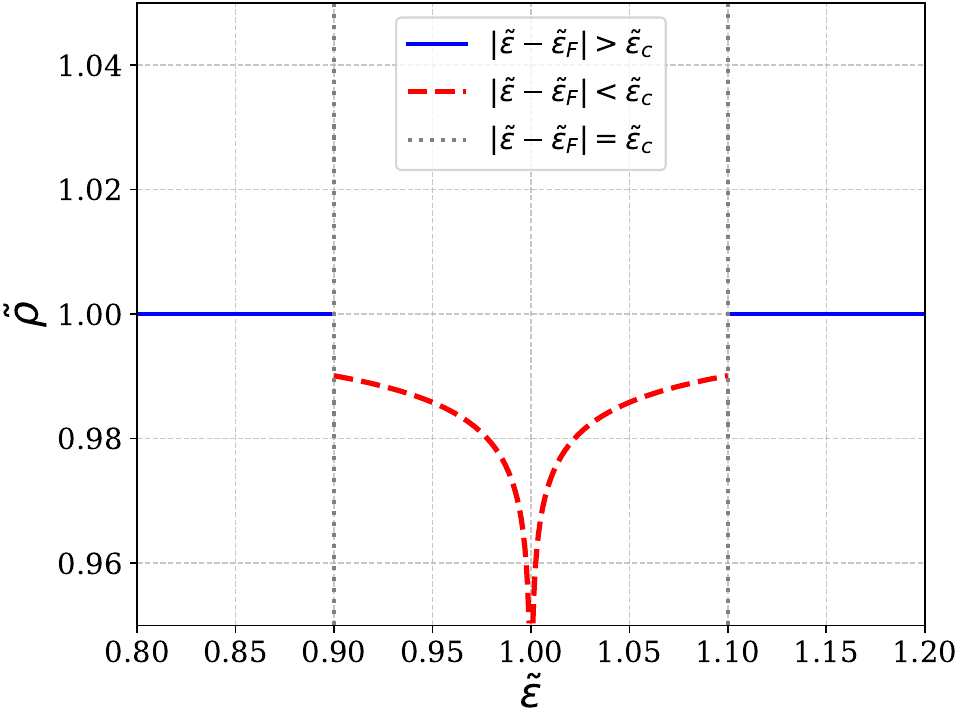}
    \caption{Dimensionless density of states per transverse conducting channel $\tilde{\rho}$ as a function of energy at the critical SOC strength $\sqrt{\tilde{\alpha}^2+\tilde{\beta}^2} = 0.57$. The Fermi energy and the critical energy are chosen as $\varepsilon_F = 1/\tau$ and $\varepsilon_c = 1/(10\tau)$.}
    \label{fig:full_DOS_cr}
\end{figure}

Determining the DOS correction allows us to calculate the renormalized single-particle DOS per transverse conducting channel, defined as
$
\rho=\rho_0+\delta\rho/N,
$
where $\rho_0=\sum_\sigma \rho_0^\sigma$ is the unperturbed DOS per transverse channel, while $\delta\rho$ denotes the interaction-induced correction obtained from Eq.~(\ref{eq:cor_dos}), in which the transverse degrees of freedom are included through the integration over $k_y$. The results for the dimensionless renormalized DOS per transverse conducting channel, $\tilde{\rho} = \rho/\rho_0$, are plotted as a function of the effective spin-orbit coupling strength, $\sqrt{\tilde{\alpha}^2 + \tilde{\beta}^2}$, in Fig.~\ref{fig:full_DOS}a for the $|\varepsilon-\varepsilon_F|<\varepsilon_c$ regime, and in Fig.~\ref{fig:full_DOS}b for the $|\varepsilon-\varepsilon_F|>\varepsilon_c$ regime, assuming $N=100$ transverse channels. As seen in Fig.~\ref{fig:full_DOS}a, within the $|\varepsilon-\varepsilon_F|<\varepsilon_c$ window, $\tilde{\rho}$ increases as the energy deviates further from the Fermi level. However, increasing the SOC strength strictly suppresses these values. Conversely, Fig.~\ref{fig:full_DOS}b demonstrates that in the quasi-1D regime, $\tilde{\rho}$ is enhanced by the SOCs. In this regime, for weak SOCs $\left(\sqrt{\tilde{\alpha}^2 + \tilde{\beta}^2} < 0.57\right)$, $\tilde{\rho}$ continues to grow as $|\varepsilon-\varepsilon_F|$ increases. Remarkably, at the critical SOC strength $\sqrt{\tilde{\alpha}^2+\tilde{\beta}^2} = 0.57$, the dimensionless DOS in the quasi-1D regime is restored to unity. This implies that the SOC perfectly compensates for the electron-electron interaction, effectively driving the correlation correction to zero.  At this critical value of the SOC strength, the variation of $\tilde{\rho}$ with $\tilde{\varepsilon}$ is shown separately in Fig.~\ref{fig:full_DOS_cr}.
For stronger SOCs $\left(\sqrt{\tilde{\alpha}^2 + \tilde{\beta}^2} > 0.57\right)$, while the overall magnitude of $\tilde{\rho}$ rises above unity (becoming a positive anomaly), the energy dependence inverts, where for a fixed $\sqrt{\tilde{\alpha}^2 + \tilde{\beta}^2}$, $\tilde{\rho}$ now decays to smaller values as $|\varepsilon-\varepsilon_F|$ increases.

\begin{figure}
    \centering
    \includegraphics[width=0.48\textwidth]{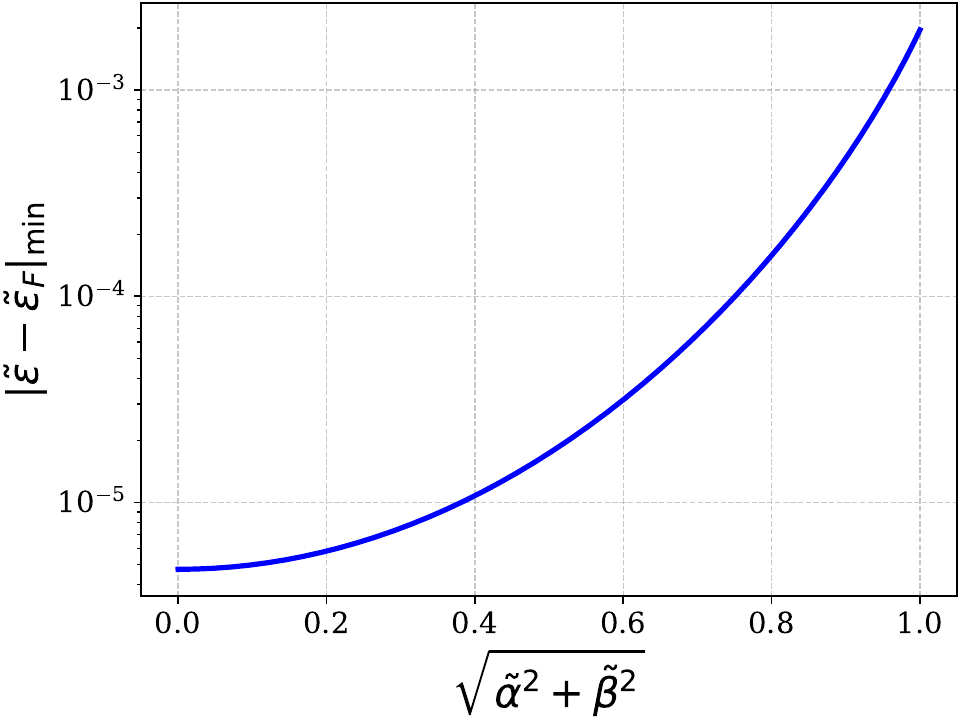}
    \caption{SOC dependence of the lower energy cutoff
$|\tilde{\varepsilon}-\tilde{\varepsilon}_F|_{\min}$ for the perturbative
validity of the DOS correction.}
    \label{fig:DOS_valid}
\end{figure}

Finally, it is important to note that the logarithmic divergence at the
Fermi level requires an explicit lower-energy cutoff for the perturbative
DOS correction. For a given SOC strength $\sqrt{\alpha^2+\beta^2}$, this cutoff is the minimum allowed distance from the Fermi level and will be denoted by
$|\varepsilon-\varepsilon_F|_{\min}$. The perturbative result is meaningful only while
the correction remains small compared with the background DOS:
\begin{equation}
    \left|
    \frac{
    \delta\rho\left(|\varepsilon-\varepsilon_F|,\sqrt{\alpha^2+\beta^2}\right)}
    {N\rho_0\left(\sqrt{\alpha^2+\beta^2}\right)}
    \right| \ll 1 .
\end{equation}
This condition is consistent with the known limitation of the
Altshuler--Aronov perturbative DOS correction, where a resummation of the
Dyson equation is required to remove the unphysical divergence when the
correction becomes large near the Fermi level~\cite{Sasioglu2009}.
For each value of $\sqrt{\alpha^2+\beta^2}$,
$|\varepsilon-\varepsilon_F|_{\min}$ is determined from
\begin{equation}
    \left|
    \frac{
    \delta\rho\left(|\varepsilon-\varepsilon_F|_{\min},
    \sqrt{\alpha^2+\beta^2}\right)}
    {N\rho_0\left(\sqrt{\alpha^2+\beta^2}\right)}
    \right| = \eta_0 ,
\end{equation}
where $\eta_0=0.1$ is used in the numerical estimate. Hence,
\begin{equation}
    |\varepsilon-\varepsilon_F|
    <
    |\varepsilon-\varepsilon_F|_{\min}
\end{equation}
is outside the perturbative domain, while the valid low-energy region is
\begin{equation}
    |\varepsilon-\varepsilon_F|_{\min}
    <
    |\varepsilon-\varepsilon_F|
    <
    \varepsilon_c .
\end{equation}
As shown in Fig.~\ref{fig:DOS_valid}, the lower cutoff
$|\tilde{\varepsilon}-\tilde{\varepsilon}_F|_{\min}$ increases monotonically
with the SOC strength $\sqrt{\tilde{\alpha}^2+\tilde{\beta}^2}$. This confirms
that stronger SOC widens the excluded energy region around the Fermi level,
where the logarithmic correction becomes non-perturbative.

\section{\label{sec:Conclusions}Conclusions}

We have analyzed the Altshuler–Aronov interaction correction to the single-particle DOS 
in a strongly anisotropic 2D metal modeled as a planar array of quasi-1D tubes with weak, 
spin-independent disorder and intrinsic Rashba/Dresselhaus SOC acting along the longitudinal 
direction. Within an exchange–only, diffusive framework, we re-summed the impurity ladder 
(charge/singlet diffuson) and used the RPA screening to obtain a compact result exhibiting a 
dimensional crossover governed by $\varepsilon_c\!\sim\! t_y^2\tau$. 
For energies well below this scale, the interaction-induced DOS anomaly has a two-dimensional, 
logarithmic singularity. For energies well above it, the anomaly crosses into the one-dimensional 
regime with the stronger square-root–type singularity.
Crucially, we have demonstrated that the impact of intrinsic SOCs strictly depends on this effective dimensionality. Close to the Fermi level, SOC deepens the two-dimensional logarithmic DOS dip by increasing the magnitude of the negative interaction correction. Further from the Fermi level, the physical role of the SOC remarkably reverses, acting instead to prominently enhance the amplitude of the sharper square-root singularity. Furthermore, we found that strong SOC in this quasi-1D regime can invert the sign of the anomaly entirely, yielding a positive DOS correction. This sign reversal fundamentally alters the energy dependence of the spectrum. Unlike the standard negative correction, which recovers at higher energies, the SOC-induced positive correction decays to smaller values as energy increases beyond $\varepsilon_F + \varepsilon_c$. Consequently, the total DOS can be tuned via SOCs from a suppressed dip to a fully restored or enhanced state. 

Our predictions are directly testable in anisotropic spin-orbit-coupled thin films such as 2D Te, where strong and gate-tunable SOC has already been established experimentally \cite{Niu2020, Geldiyev2023}. The most direct probe is low-temperature tunneling spectroscopy, since the differential conductance $dI / dV$ tracks the single-particle DOS and is known to resolve Altshuler–Aronov zero-bias anomalies \cite{Ossi2013, Liu2014}. In the present case, one should observe (i) a low-bias logarithmic DOS dip, (ii) a crossover at $eV \sim \varepsilon_c$ to a sharper quasi-1D anomaly, and (iii) a systematic SOC-driven evolution in which the low-energy dip is weakened, the finite-bias anomaly is enhanced, and both vanish at a critical SOC strength before changing sign. Observation of this sequence would provide a clear spectroscopic signature of SOC-modulated interaction effects in anisotropic diffusive conductors.

\appendix

\section{\label{sec:Appendix_A} Derivation of the Dynamically Screened Interaction}

\begin{figure}
    \centering
    \includegraphics[width=0.5\textwidth]{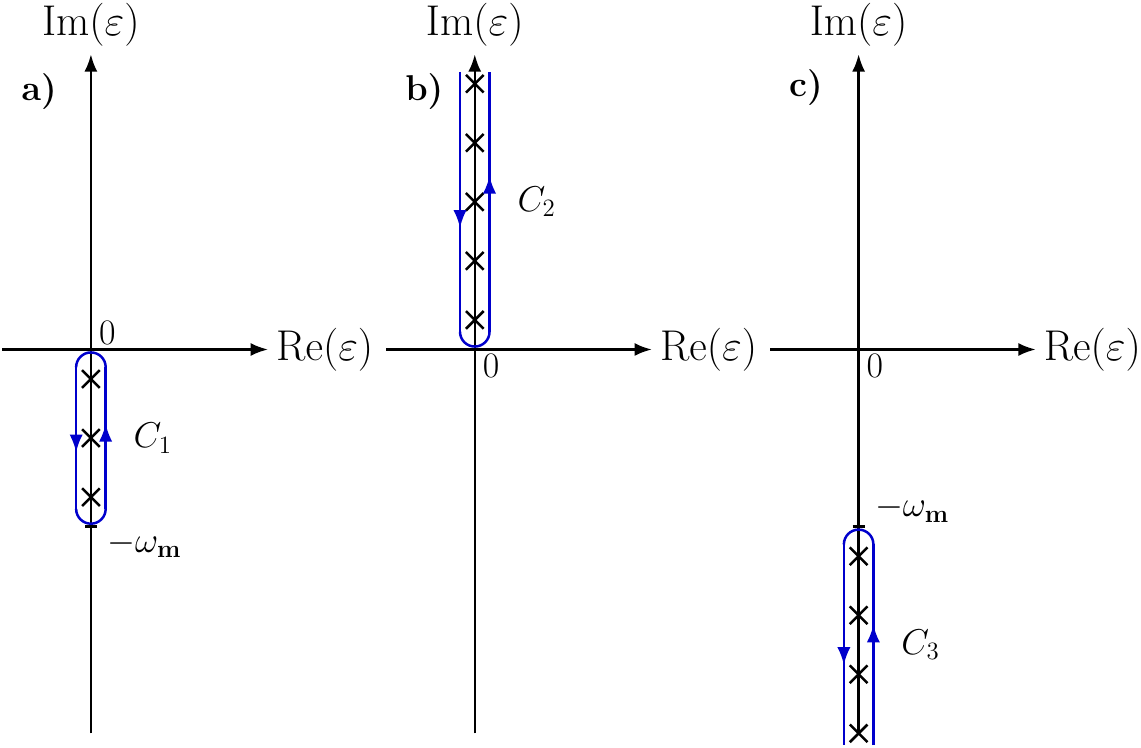}
    \caption{Integration contours in the continuous complex frequency plane used to evaluate the Matsubara sum for the polarization operator. Crosses represent discrete fermionic Matsubara frequencies $i\varepsilon_n$.}
    \label{fig:sum}
\end{figure}

In this Appendix, we detail the analytic continuation and momentum integration required to evaluate the dynamically screened interaction $V_s(\boldsymbol{q}, i\omega_m)$.
To this end, we rewrite Eq.~(\ref{eq:Vs_first}) as
\begin{equation}\label{eq:Dy_app}
    V_s(\mathbf{q}, i\omega_m) = \frac{V_0(\mathbf{q})}{1 - V_0(\mathbf{q}) \Pi_s(\mathbf{q}, i\omega_m)},
\end{equation}
where $\Pi_s(\mathbf{q}, i\omega_m)$ is defined by Eq.~(\ref{eq:polarizasyon}).
To properly evaluate Eq.~(\ref{eq:polarizasyon}), the Matsubara summation over $\varepsilon_n$ must be performed before the $\mathbf{k}$-integration to avoid discarding non-zero frequency sectors. By converting the discrete sum into a contour integral via Eq.~(\ref{eq: help_int}), the branch cuts of the Green's functions at $\mathrm{Im}(\varepsilon) = 0$ and $\mathrm{Im}(\varepsilon) = -\omega_m$ naturally decompose the polarization operator into three distinct analytic contributions, as illustrated in Fig.~\ref{fig:sum}:
\begin{equation}
    \Pi_s(\mathbf{q}, i\omega_m) = \Pi_s^{(C_1)} + \Pi_s^{(C_2)} + \Pi_s^{(C_3)}.
\end{equation}
For the outer contours $C_2$ (where $\varepsilon_n > 0$) and $C_3$ (where $\varepsilon_n < -\omega_m$), the product of the Green's functions is strictly retarded-retarded and advanced-advanced, respectively. In these frequency sectors, the elementary impurity bubbles vanish, meaning there are no vertex corrections ($\theta_s = 1$). Performing the momentum and energy integrations via the residue theorem in these unrenormalized sectors captures the bare, static compressibility of the 2D gas:
\begin{equation}
    \Pi_s^{(C_2)}(\mathbf{q}, i\omega_m) + \Pi_s^{(C_3)}(\mathbf{q}, i\omega_m) = - \sum_\sigma \rho_0^\sigma.
    \label{eq:Pi_outer}
\end{equation}

Conversely, the central contour $C_1$ strictly encloses the region $-\omega_m \le \varepsilon_n \le 0$, where the Green's functions take the retarded-advanced form. It is exclusively in this anomalous sector that the vertex acquires the macroscopic impurity-ladder correction $\theta_s \neq 1$. To evaluate $\Pi_s^{(C_1)}$, we convert the momentum integral into an integral over the energy dispersion $\xi$. Closing the contour in the upper half-plane encloses the single pole of the retarded Green's function at $\xi_{\text{pole}} = \varepsilon + \omega_m - \Delta\xi^\sigma_{\mathbf{q}} + \frac{i}{2\tau}$. Evaluating this via the residue theorem, applying angular averaging over the Fermi surface, and dressing the result with the diffusive ladder $\theta_s$, we obtain the dynamic correction:
\begin{equation}
    \Pi_s^{(C_1)}(\mathbf{q}, i\omega_m) = \sum_\sigma \rho_0^\sigma \frac{|\omega_m|}{|\omega_m| + V_1^\sigma(\mathbf{q})},
    \label{eq:Pi_inner}
\end{equation}
where $V_1^\sigma(\mathbf{q})$ is the anisotropic diffusion operator defined in Eq.~(\ref{eq:V1}).
Summing all three contour contributions (Eqs.~(\ref{eq:Pi_outer}) and (\ref{eq:Pi_inner})) yields the exact, total polarization operator:
\begin{eqnarray}
    \Pi_s(\mathbf{q}, i\omega_m) &=& - \sum_\sigma \rho_0^\sigma + \sum_\sigma \rho_0^\sigma \frac{|\omega_m|}{|\omega_m| + V_1^\sigma(\mathbf{q})} \nonumber \\
    &=& - \sum_\sigma \rho_0^\sigma \frac{V_1^\sigma(\mathbf{q})}{|\omega_m| + V_1^\sigma(\mathbf{q})}.
\end{eqnarray}
Finally, substituting this dressed polarization operator and the bare Coulomb kernel $V_0(\mathbf{q}) = 2\pi e^2 / |\mathbf{q}|$ into the Dyson equation (\ref{eq:Dy_app}), we divide the numerator and denominator by $V_0(\mathbf{q})$ to obtain:
\begin{equation}
    V_s(\mathbf{q}, i\omega_m) = \left( \frac{|\mathbf{q}|}{2\pi e^2} + \sum_\sigma \rho_0^\sigma \frac{V_1^\sigma(\mathbf{q})}{|\omega_m| + V_1^\sigma(\mathbf{q})} \right)^{-1}.
\end{equation}
Factoring out $(2\pi e^2)^{-1}$ immediately recovers Eq.~(\ref{eq:V_main}) of the main text, completing the derivation.

\begin{acknowledgments}
BT acknowledges support from the Turkish Academy of Sciences (TUBA) under Grant No. AD-2026.
\end{acknowledgments}

\bibliography{SOC}

\providecommand{\noopsort}[1]{}\providecommand{\singleletter}[1]{#1}%
\begin{thebibliography}{57}%
\makeatletter
\providecommand \@ifxundefined [1]{%
 \@ifx{#1\undefined}
}%
\providecommand \@ifnum [1]{%
 \ifnum #1\expandafter \@firstoftwo
 \else \expandafter \@secondoftwo
 \fi
}%
\providecommand \@ifx [1]{%
 \ifx #1\expandafter \@firstoftwo
 \else \expandafter \@secondoftwo
 \fi
}%
\providecommand \natexlab [1]{#1}%
\providecommand \enquote  [1]{``#1''}%
\providecommand \bibnamefont  [1]{#1}%
\providecommand \bibfnamefont [1]{#1}%
\providecommand \citenamefont [1]{#1}%
\providecommand \href@noop [0]{\@secondoftwo}%
\providecommand \href [0]{\begingroup \@sanitize@url \@href}%
\providecommand \@href[1]{\@@startlink{#1}\@@href}%
\providecommand \@@href[1]{\endgroup#1\@@endlink}%
\providecommand \@sanitize@url [0]{\catcode `\\12\catcode `\$12\catcode
  `\&12\catcode `\#12\catcode `\^12\catcode `\_12\catcode `\%12\relax}%
\providecommand \@@startlink[1]{}%
\providecommand \@@endlink[0]{}%
\providecommand \url  [0]{\begingroup\@sanitize@url \@url }%
\providecommand \@url [1]{\endgroup\@href {#1}{\urlprefix }}%
\providecommand \urlprefix  [0]{URL }%
\providecommand \Eprint [0]{\href }%
\providecommand \doibase [0]{https://doi.org/}%
\providecommand \selectlanguage [0]{\@gobble}%
\providecommand \bibinfo  [0]{\@secondoftwo}%
\providecommand \bibfield  [0]{\@secondoftwo}%
\providecommand \translation [1]{[#1]}%
\providecommand \BibitemOpen [0]{}%
\providecommand \bibitemStop [0]{}%
\providecommand \bibitemNoStop [0]{.\EOS\space}%
\providecommand \EOS [0]{\spacefactor3000\relax}%
\providecommand \BibitemShut  [1]{\csname bibitem#1\endcsname}%
\let\auto@bib@innerbib\@empty
\bibitem [{\citenamefont {Miller}\ \emph {et~al.}(2003)\citenamefont {Miller},
  \citenamefont {Zumb\"uhl}, \citenamefont {Marcus}, \citenamefont
  {Lyanda-Geller}, \citenamefont {Goldhaber-Gordon}, \citenamefont {Campman},\
  and\ \citenamefont {Gossard}}]{Miller2003}%
  \BibitemOpen
  \bibfield  {author} {\bibinfo {author} {\bibfnamefont {J.~B.}\ \bibnamefont
  {Miller}}, \bibinfo {author} {\bibfnamefont {D.~M.}\ \bibnamefont
  {Zumb\"uhl}}, \bibinfo {author} {\bibfnamefont {C.~M.}\ \bibnamefont
  {Marcus}}, \bibinfo {author} {\bibfnamefont {Y.~B.}\ \bibnamefont
  {Lyanda-Geller}}, \bibinfo {author} {\bibfnamefont {D.}~\bibnamefont
  {Goldhaber-Gordon}}, \bibinfo {author} {\bibfnamefont {K.}~\bibnamefont
  {Campman}},\ and\ \bibinfo {author} {\bibfnamefont {A.~C.}\ \bibnamefont
  {Gossard}},\ }\bibfield  {title} {\bibinfo {title} {{Gate-Controlled
  Spin-Orbit Quantum Interference Effects in Lateral Transport}},\ }\href
  {https://doi.org/10.1103/PhysRevLett.90.076807} {\bibfield  {journal}
  {\bibinfo  {journal} {Phys. Rev. Lett.}\ }\textbf {\bibinfo {volume} {90}},\
  \bibinfo {pages} {076807} (\bibinfo {year} {2003})}\BibitemShut {NoStop}%
\bibitem [{\citenamefont {Caviglia}\ \emph {et~al.}(2010)\citenamefont
  {Caviglia}, \citenamefont {Gabay}, \citenamefont {Gariglio}, \citenamefont
  {Reyren}, \citenamefont {Cancellieri},\ and\ \citenamefont
  {Triscone}}]{Caviglia2010}%
  \BibitemOpen
  \bibfield  {author} {\bibinfo {author} {\bibfnamefont {A.~D.}\ \bibnamefont
  {Caviglia}}, \bibinfo {author} {\bibfnamefont {M.}~\bibnamefont {Gabay}},
  \bibinfo {author} {\bibfnamefont {S.}~\bibnamefont {Gariglio}}, \bibinfo
  {author} {\bibfnamefont {N.}~\bibnamefont {Reyren}}, \bibinfo {author}
  {\bibfnamefont {C.}~\bibnamefont {Cancellieri}},\ and\ \bibinfo {author}
  {\bibfnamefont {J.-M.}\ \bibnamefont {Triscone}},\ }\bibfield  {title}
  {\bibinfo {title} {{Tunable Rashba Spin-Orbit Interaction at Oxide
  Interfaces}},\ }\href {https://doi.org/10.1103/PhysRevLett.104.126803}
  {\bibfield  {journal} {\bibinfo  {journal} {Phys. Rev. Lett.}\ }\textbf
  {\bibinfo {volume} {104}},\ \bibinfo {pages} {126803} (\bibinfo {year}
  {2010})}\BibitemShut {NoStop}%
\bibitem [{\citenamefont {Guo}\ \emph {et~al.}(2021)\citenamefont {Guo},
  \citenamefont {Yan}, \citenamefont {Xu}, \citenamefont {Li},\ and\
  \citenamefont {Zeng}}]{Guo2021}%
  \BibitemOpen
  \bibfield  {author} {\bibinfo {author} {\bibfnamefont {L.}~\bibnamefont
  {Guo}}, \bibinfo {author} {\bibfnamefont {Y.}~\bibnamefont {Yan}}, \bibinfo
  {author} {\bibfnamefont {R.}~\bibnamefont {Xu}}, \bibinfo {author}
  {\bibfnamefont {J.}~\bibnamefont {Li}},\ and\ \bibinfo {author}
  {\bibfnamefont {C.}~\bibnamefont {Zeng}},\ }\bibfield  {title} {\bibinfo
  {title} {{Zero-Bias Conductance Peaks Effectively Tuned by Gating-Controlled
  Rashba Spin-Orbit Coupling}},\ }\href
  {https://doi.org/10.1103/PhysRevLett.126.057701} {\bibfield  {journal}
  {\bibinfo  {journal} {Phys. Rev. Lett.}\ }\textbf {\bibinfo {volume} {126}},\
  \bibinfo {pages} {057701} (\bibinfo {year} {2021})}\BibitemShut {NoStop}%
\bibitem [{\citenamefont {Mariani}\ \emph {et~al.}(2007)\citenamefont
  {Mariani}, \citenamefont {Glazman}, \citenamefont {Kamenev},\ and\
  \citenamefont {von Oppen}}]{Mariani2007}%
  \BibitemOpen
  \bibfield  {author} {\bibinfo {author} {\bibfnamefont {E.}~\bibnamefont
  {Mariani}}, \bibinfo {author} {\bibfnamefont {L.~I.}\ \bibnamefont
  {Glazman}}, \bibinfo {author} {\bibfnamefont {A.}~\bibnamefont {Kamenev}},\
  and\ \bibinfo {author} {\bibfnamefont {F.}~\bibnamefont {von Oppen}},\
  }\bibfield  {title} {\bibinfo {title} {Zero-bias anomaly in the tunneling
  density of states of graphene},\ }\href
  {https://doi.org/10.1103/PhysRevB.76.165402} {\bibfield  {journal} {\bibinfo
  {journal} {Phys. Rev. B}\ }\textbf {\bibinfo {volume} {76}},\ \bibinfo
  {pages} {165402} (\bibinfo {year} {2007})}\BibitemShut {NoStop}%
\bibitem [{\citenamefont {{Wang}}\ \emph {et~al.}(2015)\citenamefont {{Wang}},
  \citenamefont {{Ki}}, \citenamefont {{Chen}}, \citenamefont {{Berger}},
  \citenamefont {{MacDonald}},\ and\ \citenamefont {{Morpurgo}}}]{Wang2015}%
  \BibitemOpen
  \bibfield  {author} {\bibinfo {author} {\bibfnamefont {Z.}~\bibnamefont
  {{Wang}}}, \bibinfo {author} {\bibfnamefont {D.-K.}\ \bibnamefont {{Ki}}},
  \bibinfo {author} {\bibfnamefont {H.}~\bibnamefont {{Chen}}}, \bibinfo
  {author} {\bibfnamefont {H.}~\bibnamefont {{Berger}}}, \bibinfo {author}
  {\bibfnamefont {A.~H.}\ \bibnamefont {{MacDonald}}},\ and\ \bibinfo {author}
  {\bibfnamefont {A.~F.}\ \bibnamefont {{Morpurgo}}},\ }\bibfield  {title}
  {\bibinfo {title} {{Strong interface-induced spin-orbit interaction in
  graphene on WS$_{2}$}},\ }\href {https://doi.org/10.1038/ncomms9339}
  {\bibfield  {journal} {\bibinfo  {journal} {Nature Communications}\ }\textbf
  {\bibinfo {volume} {6}},\ \bibinfo {eid} {8339} (\bibinfo {year}
  {2015})}\BibitemShut {NoStop}%
\bibitem [{\citenamefont {Wakamura}\ \emph {et~al.}(2018)\citenamefont
  {Wakamura}, \citenamefont {Reale}, \citenamefont {Palczynski}, \citenamefont
  {Gu\'eron}, \citenamefont {Mattevi},\ and\ \citenamefont
  {Bouchiat}}]{Wakamura2018}%
  \BibitemOpen
  \bibfield  {author} {\bibinfo {author} {\bibfnamefont {T.}~\bibnamefont
  {Wakamura}}, \bibinfo {author} {\bibfnamefont {F.}~\bibnamefont {Reale}},
  \bibinfo {author} {\bibfnamefont {P.}~\bibnamefont {Palczynski}}, \bibinfo
  {author} {\bibfnamefont {S.}~\bibnamefont {Gu\'eron}}, \bibinfo {author}
  {\bibfnamefont {C.}~\bibnamefont {Mattevi}},\ and\ \bibinfo {author}
  {\bibfnamefont {H.}~\bibnamefont {Bouchiat}},\ }\bibfield  {title} {\bibinfo
  {title} {{Strong Anisotropic Spin-Orbit Interaction Induced in Graphene by
  Monolayer ${\mathrm{WS}}_{2}$}},\ }\href
  {https://doi.org/10.1103/PhysRevLett.120.106802} {\bibfield  {journal}
  {\bibinfo  {journal} {Phys. Rev. Lett.}\ }\textbf {\bibinfo {volume} {120}},\
  \bibinfo {pages} {106802} (\bibinfo {year} {2018})}\BibitemShut {NoStop}%
\bibitem [{\citenamefont {{F{\"u}l{\"o}p}}\ \emph {et~al.}(2021)\citenamefont
  {{F{\"u}l{\"o}p}}, \citenamefont {{M{\'a}rffy}}, \citenamefont {{Zihlmann}},
  \citenamefont {{Gmitra}}, \citenamefont {{T{\'o}v{\'a}ri}}, \citenamefont
  {{Szentp{\'e}teri}}, \citenamefont {{Kedves}}, \citenamefont {{Watanabe}},
  \citenamefont {{Taniguchi}}, \citenamefont {{Fabian}}, \citenamefont
  {{Sch{\"o}nenberger}}, \citenamefont {{Makk}},\ and\ \citenamefont
  {{Csonka}}}]{Albin2021}%
  \BibitemOpen
  \bibfield  {author} {\bibinfo {author} {\bibfnamefont {B.}~\bibnamefont
  {{F{\"u}l{\"o}p}}}, \bibinfo {author} {\bibfnamefont {A.}~\bibnamefont
  {{M{\'a}rffy}}}, \bibinfo {author} {\bibfnamefont {S.}~\bibnamefont
  {{Zihlmann}}}, \bibinfo {author} {\bibfnamefont {M.}~\bibnamefont
  {{Gmitra}}}, \bibinfo {author} {\bibfnamefont {E.}~\bibnamefont
  {{T{\'o}v{\'a}ri}}}, \bibinfo {author} {\bibfnamefont {B.}~\bibnamefont
  {{Szentp{\'e}teri}}}, \bibinfo {author} {\bibfnamefont {M.}~\bibnamefont
  {{Kedves}}}, \bibinfo {author} {\bibfnamefont {K.}~\bibnamefont
  {{Watanabe}}}, \bibinfo {author} {\bibfnamefont {T.}~\bibnamefont
  {{Taniguchi}}}, \bibinfo {author} {\bibfnamefont {J.}~\bibnamefont
  {{Fabian}}}, \bibinfo {author} {\bibfnamefont {C.}~\bibnamefont
  {{Sch{\"o}nenberger}}}, \bibinfo {author} {\bibfnamefont {P.}~\bibnamefont
  {{Makk}}},\ and\ \bibinfo {author} {\bibfnamefont {S.}~\bibnamefont
  {{Csonka}}},\ }\bibfield  {title} {\bibinfo {title} {{Boosting proximity spin
  orbit coupling in graphene/WSe$_2$ heterostructures via hydrostatic
  pressure}},\ }\href {https://doi.org/10.1038/s41699-021-00262-9} {\bibfield
  {journal} {\bibinfo  {journal} {npj 2d materials and applications}\ }\textbf
  {\bibinfo {volume} {5}},\ \bibinfo {pages} {82} (\bibinfo {year}
  {2021})}\BibitemShut {NoStop}%
\bibitem [{\citenamefont {{Sun}}\ \emph {et~al.}(2023)\citenamefont {{Sun}},
  \citenamefont {{Rademaker}}, \citenamefont {{Mauro}}, \citenamefont
  {{Scarfato}}, \citenamefont {{P{\'a}sztor}}, \citenamefont
  {{Guti{\'e}rrez-Lezama}}, \citenamefont {{Wang}}, \citenamefont
  {{Martinez-Castro}}, \citenamefont {{Morpurgo}},\ and\ \citenamefont
  {{Renner}}}]{Sun2023}%
  \BibitemOpen
  \bibfield  {author} {\bibinfo {author} {\bibfnamefont {L.}~\bibnamefont
  {{Sun}}}, \bibinfo {author} {\bibfnamefont {L.}~\bibnamefont {{Rademaker}}},
  \bibinfo {author} {\bibfnamefont {D.}~\bibnamefont {{Mauro}}}, \bibinfo
  {author} {\bibfnamefont {A.}~\bibnamefont {{Scarfato}}}, \bibinfo {author}
  {\bibfnamefont {{\'A}.}~\bibnamefont {{P{\'a}sztor}}}, \bibinfo {author}
  {\bibfnamefont {I.}~\bibnamefont {{Guti{\'e}rrez-Lezama}}}, \bibinfo {author}
  {\bibfnamefont {Z.}~\bibnamefont {{Wang}}}, \bibinfo {author} {\bibfnamefont
  {J.}~\bibnamefont {{Martinez-Castro}}}, \bibinfo {author} {\bibfnamefont
  {A.~F.}\ \bibnamefont {{Morpurgo}}},\ and\ \bibinfo {author} {\bibfnamefont
  {C.}~\bibnamefont {{Renner}}},\ }\bibfield  {title} {\bibinfo {title}
  {{Determining spin-orbit coupling in graphene by quasiparticle interference
  imaging}},\ }\href {https://doi.org/10.1038/s41467-023-39453-x} {\bibfield
  {journal} {\bibinfo  {journal} {Nature Communications}\ }\textbf {\bibinfo
  {volume} {14}},\ \bibinfo {eid} {3771} (\bibinfo {year} {2023})}\BibitemShut
  {NoStop}%
\bibitem [{\citenamefont {Endo}\ and\ \citenamefont {Iye}(2005)}]{Endo2005}%
  \BibitemOpen
  \bibfield  {author} {\bibinfo {author} {\bibfnamefont {A.}~\bibnamefont
  {Endo}}\ and\ \bibinfo {author} {\bibfnamefont {Y.}~\bibnamefont {Iye}},\
  }\bibfield  {title} {\bibinfo {title} {Origin of positive magnetoresistance
  in small-amplitude unidirectional lateral superlattices},\ }\href
  {https://doi.org/10.1103/PhysRevB.72.235303} {\bibfield  {journal} {\bibinfo
  {journal} {Phys. Rev. B}\ }\textbf {\bibinfo {volume} {72}},\ \bibinfo
  {pages} {235303} (\bibinfo {year} {2005})}\BibitemShut {NoStop}%
\bibitem [{\citenamefont {Endo}\ \emph {et~al.}(2021)\citenamefont {Endo},
  \citenamefont {Katsumoto},\ and\ \citenamefont {Iye}}]{Endo2021}%
  \BibitemOpen
  \bibfield  {author} {\bibinfo {author} {\bibfnamefont {A.}~\bibnamefont
  {Endo}}, \bibinfo {author} {\bibfnamefont {S.}~\bibnamefont {Katsumoto}},\
  and\ \bibinfo {author} {\bibfnamefont {Y.}~\bibnamefont {Iye}},\ }\bibfield
  {title} {\bibinfo {title} {{Commensurability oscillations in the Hall
  resistance of unidirectional lateral superlattices}},\ }\href
  {https://doi.org/10.1103/PhysRevB.103.235303} {\bibfield  {journal} {\bibinfo
   {journal} {Phys. Rev. B}\ }\textbf {\bibinfo {volume} {103}},\ \bibinfo
  {pages} {235303} (\bibinfo {year} {2021})}\BibitemShut {NoStop}%
\bibitem [{\citenamefont {{Blumenstein}}\ \emph {et~al.}(2011)\citenamefont
  {{Blumenstein}}, \citenamefont {{Sch{\"a}fer}}, \citenamefont {{Mietke}},
  \citenamefont {{Meyer}}, \citenamefont {{Dollinger}}, \citenamefont
  {{Lochner}}, \citenamefont {{Cui}}, \citenamefont {{Patthey}}, \citenamefont
  {{Matzdorf}},\ and\ \citenamefont {{Claessen}}}]{Blumenstein2011}%
  \BibitemOpen
  \bibfield  {author} {\bibinfo {author} {\bibfnamefont {C.}~\bibnamefont
  {{Blumenstein}}}, \bibinfo {author} {\bibfnamefont {J.}~\bibnamefont
  {{Sch{\"a}fer}}}, \bibinfo {author} {\bibfnamefont {S.}~\bibnamefont
  {{Mietke}}}, \bibinfo {author} {\bibfnamefont {S.}~\bibnamefont {{Meyer}}},
  \bibinfo {author} {\bibfnamefont {A.}~\bibnamefont {{Dollinger}}}, \bibinfo
  {author} {\bibfnamefont {M.}~\bibnamefont {{Lochner}}}, \bibinfo {author}
  {\bibfnamefont {X.~Y.}\ \bibnamefont {{Cui}}}, \bibinfo {author}
  {\bibfnamefont {L.}~\bibnamefont {{Patthey}}}, \bibinfo {author}
  {\bibfnamefont {R.}~\bibnamefont {{Matzdorf}}},\ and\ \bibinfo {author}
  {\bibfnamefont {R.}~\bibnamefont {{Claessen}}},\ }\bibfield  {title}
  {\bibinfo {title} {{Atomically controlled quantum chains hosting a
  Tomonaga-Luttinger liquid}},\ }\href {https://doi.org/10.1038/nphys2051}
  {\bibfield  {journal} {\bibinfo  {journal} {Nature Physics}\ }\textbf
  {\bibinfo {volume} {7}},\ \bibinfo {pages} {776} (\bibinfo {year}
  {2011})}\BibitemShut {NoStop}%
\bibitem [{\citenamefont {Tegenkamp}\ \emph {et~al.}(2012)\citenamefont
  {Tegenkamp}, \citenamefont {L\"ukermann}, \citenamefont {Pfn\"ur},
  \citenamefont {Slomski}, \citenamefont {Landolt},\ and\ \citenamefont
  {Dil}}]{Tegenkamp2012}%
  \BibitemOpen
  \bibfield  {author} {\bibinfo {author} {\bibfnamefont {C.}~\bibnamefont
  {Tegenkamp}}, \bibinfo {author} {\bibfnamefont {D.}~\bibnamefont
  {L\"ukermann}}, \bibinfo {author} {\bibfnamefont {H.}~\bibnamefont
  {Pfn\"ur}}, \bibinfo {author} {\bibfnamefont {B.}~\bibnamefont {Slomski}},
  \bibinfo {author} {\bibfnamefont {G.}~\bibnamefont {Landolt}},\ and\ \bibinfo
  {author} {\bibfnamefont {J.~H.}\ \bibnamefont {Dil}},\ }\bibfield  {title}
  {\bibinfo {title} {{Fermi Nesting between Atomic Wires with Strong Spin-Orbit
  Coupling}},\ }\href {https://doi.org/10.1103/PhysRevLett.109.266401}
  {\bibfield  {journal} {\bibinfo  {journal} {Phys. Rev. Lett.}\ }\textbf
  {\bibinfo {volume} {109}},\ \bibinfo {pages} {266401} (\bibinfo {year}
  {2012})}\BibitemShut {NoStop}%
\bibitem [{\citenamefont {Park}\ \emph {et~al.}(2013)\citenamefont {Park},
  \citenamefont {Jung}, \citenamefont {Jung}, \citenamefont {Yamane},
  \citenamefont {Kosugi},\ and\ \citenamefont {Yeom}}]{Park2013}%
  \BibitemOpen
  \bibfield  {author} {\bibinfo {author} {\bibfnamefont {J.}~\bibnamefont
  {Park}}, \bibinfo {author} {\bibfnamefont {S.~W.}\ \bibnamefont {Jung}},
  \bibinfo {author} {\bibfnamefont {M.-C.}\ \bibnamefont {Jung}}, \bibinfo
  {author} {\bibfnamefont {H.}~\bibnamefont {Yamane}}, \bibinfo {author}
  {\bibfnamefont {N.}~\bibnamefont {Kosugi}},\ and\ \bibinfo {author}
  {\bibfnamefont {H.~W.}\ \bibnamefont {Yeom}},\ }\bibfield  {title} {\bibinfo
  {title} {{Self-Assembled Nanowires with Giant Rashba Split Bands}},\ }\href
  {https://doi.org/10.1103/PhysRevLett.110.036801} {\bibfield  {journal}
  {\bibinfo  {journal} {Phys. Rev. Lett.}\ }\textbf {\bibinfo {volume} {110}},\
  \bibinfo {pages} {036801} (\bibinfo {year} {2013})}\BibitemShut {NoStop}%
\bibitem [{\citenamefont {Ossi}\ \emph {et~al.}(2013)\citenamefont {Ossi},
  \citenamefont {Bitton}, \citenamefont {Gutman},\ and\ \citenamefont
  {Frydman}}]{Ossi2013}%
  \BibitemOpen
  \bibfield  {author} {\bibinfo {author} {\bibfnamefont {N.}~\bibnamefont
  {Ossi}}, \bibinfo {author} {\bibfnamefont {L.}~\bibnamefont {Bitton}},
  \bibinfo {author} {\bibfnamefont {D.~B.}\ \bibnamefont {Gutman}},\ and\
  \bibinfo {author} {\bibfnamefont {A.}~\bibnamefont {Frydman}},\ }\bibfield
  {title} {\bibinfo {title} {Zero-bias anomaly in a two-dimensional granular
  insulator},\ }\href {https://doi.org/10.1103/PhysRevB.87.115137} {\bibfield
  {journal} {\bibinfo  {journal} {Phys. Rev. B}\ }\textbf {\bibinfo {volume}
  {87}},\ \bibinfo {pages} {115137} (\bibinfo {year} {2013})}\BibitemShut
  {NoStop}%
\bibitem [{\citenamefont {Ovadyahu}(2022)}]{Ovadyahu2022}%
  \BibitemOpen
  \bibfield  {author} {\bibinfo {author} {\bibfnamefont {Z.}~\bibnamefont
  {Ovadyahu}},\ }\bibfield  {title} {\bibinfo {title} {Interaction-induced
  spatial correlations in a disordered glass},\ }\href
  {https://doi.org/10.1103/PhysRevB.105.235101} {\bibfield  {journal} {\bibinfo
   {journal} {Phys. Rev. B}\ }\textbf {\bibinfo {volume} {105}},\ \bibinfo
  {pages} {235101} (\bibinfo {year} {2022})}\BibitemShut {NoStop}%
\bibitem [{\citenamefont {{Altshuler}}\ and\ \citenamefont
  {{Aronov}}(1979)}]{Altshuler1979}%
  \BibitemOpen
  \bibfield  {author} {\bibinfo {author} {\bibfnamefont {B.~L.}\ \bibnamefont
  {{Altshuler}}}\ and\ \bibinfo {author} {\bibfnamefont {A.~G.}\ \bibnamefont
  {{Aronov}}},\ }\bibfield  {title} {\bibinfo {title} {{Zero bias anomaly in
  tunnel resistance and electron-electron interaction}},\ }\href
  {https://doi.org/10.1016/0038-1098(79)90967-0} {\bibfield  {journal}
  {\bibinfo  {journal} {Solid State Communications}\ }\textbf {\bibinfo
  {volume} {30}},\ \bibinfo {pages} {115} (\bibinfo {year} {1979})}\BibitemShut
  {NoStop}%
\bibitem [{\citenamefont {{Efros}}\ and\ \citenamefont
  {{Shklovskii}}(1975)}]{Efros1975}%
  \BibitemOpen
  \bibfield  {author} {\bibinfo {author} {\bibfnamefont {A.~L.}\ \bibnamefont
  {{Efros}}}\ and\ \bibinfo {author} {\bibfnamefont {B.~I.}\ \bibnamefont
  {{Shklovskii}}},\ }\bibfield  {title} {\bibinfo {title} {{Coulomb gap and low
  temperature conductivity of disordered systems}},\ }\href
  {https://doi.org/10.1088/0022-3719/8/4/003} {\bibfield  {journal} {\bibinfo
  {journal} {Journal of Physics C Solid State Physics}\ }\textbf {\bibinfo
  {volume} {8}},\ \bibinfo {pages} {L49} (\bibinfo {year} {1975})}\BibitemShut
  {NoStop}%
\bibitem [{\citenamefont {{Bockrath}}\ \emph {et~al.}(1999)\citenamefont
  {{Bockrath}}, \citenamefont {{Cobden}}, \citenamefont {{Lu}}, \citenamefont
  {{Rinzler}}, \citenamefont {{Smalley}}, \citenamefont {{Balents}},\ and\
  \citenamefont {{McEuen}}}]{Bockrath1999}%
  \BibitemOpen
  \bibfield  {author} {\bibinfo {author} {\bibfnamefont {M.}~\bibnamefont
  {{Bockrath}}}, \bibinfo {author} {\bibfnamefont {D.~H.}\ \bibnamefont
  {{Cobden}}}, \bibinfo {author} {\bibfnamefont {J.}~\bibnamefont {{Lu}}},
  \bibinfo {author} {\bibfnamefont {A.~G.}\ \bibnamefont {{Rinzler}}}, \bibinfo
  {author} {\bibfnamefont {R.~E.}\ \bibnamefont {{Smalley}}}, \bibinfo {author}
  {\bibfnamefont {L.}~\bibnamefont {{Balents}}},\ and\ \bibinfo {author}
  {\bibfnamefont {P.~L.}\ \bibnamefont {{McEuen}}},\ }\bibfield  {title}
  {\bibinfo {title} {{Luttinger-liquid behaviour in carbon nanotubes}},\ }\href
  {https://doi.org/10.1038/17569} {\bibfield  {journal} {\bibinfo  {journal}
  {\nat}\ }\textbf {\bibinfo {volume} {397}},\ \bibinfo {pages} {598} (\bibinfo
  {year} {1999})}\BibitemShut {NoStop}%
\bibitem [{\citenamefont {Turco}\ \emph {et~al.}(2024)\citenamefont {Turco},
  \citenamefont {Aapro}, \citenamefont {Ganguli}, \citenamefont {Krane},
  \citenamefont {Drost}, \citenamefont {Sobrino}, \citenamefont {Bernhardt},
  \citenamefont {Jur\'{\i}\ifmmode~\check{c}\else \v{c}\fi{}ek}, \citenamefont
  {Fasel}, \citenamefont {Ruffieux}, \citenamefont {Liljeroth},\ and\
  \citenamefont {Jacob}}]{Turco2024}%
  \BibitemOpen
  \bibfield  {author} {\bibinfo {author} {\bibfnamefont {E.}~\bibnamefont
  {Turco}}, \bibinfo {author} {\bibfnamefont {M.}~\bibnamefont {Aapro}},
  \bibinfo {author} {\bibfnamefont {S.~C.}\ \bibnamefont {Ganguli}}, \bibinfo
  {author} {\bibfnamefont {N.}~\bibnamefont {Krane}}, \bibinfo {author}
  {\bibfnamefont {R.}~\bibnamefont {Drost}}, \bibinfo {author} {\bibfnamefont
  {N.}~\bibnamefont {Sobrino}}, \bibinfo {author} {\bibfnamefont
  {A.}~\bibnamefont {Bernhardt}}, \bibinfo {author} {\bibfnamefont
  {M.}~\bibnamefont {Jur\'{\i}\ifmmode~\check{c}\else \v{c}\fi{}ek}}, \bibinfo
  {author} {\bibfnamefont {R.}~\bibnamefont {Fasel}}, \bibinfo {author}
  {\bibfnamefont {P.}~\bibnamefont {Ruffieux}}, \bibinfo {author}
  {\bibfnamefont {P.}~\bibnamefont {Liljeroth}},\ and\ \bibinfo {author}
  {\bibfnamefont {D.}~\bibnamefont {Jacob}},\ }\bibfield  {title} {\bibinfo
  {title} {{Demonstrating Kondo behavior by temperature-dependent scanning
  tunneling spectroscopy}},\ }\href
  {https://doi.org/10.1103/PhysRevResearch.6.L022061} {\bibfield  {journal}
  {\bibinfo  {journal} {Phys. Rev. Res.}\ }\textbf {\bibinfo {volume} {6}},\
  \bibinfo {pages} {L022061} (\bibinfo {year} {2024})}\BibitemShut {NoStop}%
\bibitem [{\citenamefont {Pixley}\ \emph {et~al.}(2016)\citenamefont {Pixley},
  \citenamefont {Huse},\ and\ \citenamefont {Das~Sarma}}]{Pixley2016}%
  \BibitemOpen
  \bibfield  {author} {\bibinfo {author} {\bibfnamefont {J.~H.}\ \bibnamefont
  {Pixley}}, \bibinfo {author} {\bibfnamefont {D.~A.}\ \bibnamefont {Huse}},\
  and\ \bibinfo {author} {\bibfnamefont {S.}~\bibnamefont {Das~Sarma}},\
  }\bibfield  {title} {\bibinfo {title} {{Rare-Region-Induced Avoided Quantum
  Criticality in Disordered Three-Dimensional Dirac and Weyl Semimetals}},\
  }\href {https://doi.org/10.1103/PhysRevX.6.021042} {\bibfield  {journal}
  {\bibinfo  {journal} {Phys. Rev. X}\ }\textbf {\bibinfo {volume} {6}},\
  \bibinfo {pages} {021042} (\bibinfo {year} {2016})}\BibitemShut {NoStop}%
\bibitem [{\citenamefont {Li}\ \emph {et~al.}(2020)\citenamefont {Li},
  \citenamefont {Hanus}, \citenamefont {Polanco}, \citenamefont {Zeidler},
  \citenamefont {Koblm\"uller}, \citenamefont {Koh},\ and\ \citenamefont
  {Lindsay}}]{Li2020}%
  \BibitemOpen
  \bibfield  {author} {\bibinfo {author} {\bibfnamefont {H.}~\bibnamefont
  {Li}}, \bibinfo {author} {\bibfnamefont {R.}~\bibnamefont {Hanus}}, \bibinfo
  {author} {\bibfnamefont {C.~A.}\ \bibnamefont {Polanco}}, \bibinfo {author}
  {\bibfnamefont {A.}~\bibnamefont {Zeidler}}, \bibinfo {author} {\bibfnamefont
  {G.}~\bibnamefont {Koblm\"uller}}, \bibinfo {author} {\bibfnamefont {Y.~K.}\
  \bibnamefont {Koh}},\ and\ \bibinfo {author} {\bibfnamefont {L.}~\bibnamefont
  {Lindsay}},\ }\bibfield  {title} {\bibinfo {title} {{GaN thermal transport
  limited by the interplay of dislocations and size effects}},\ }\href
  {https://doi.org/10.1103/PhysRevB.102.014313} {\bibfield  {journal} {\bibinfo
   {journal} {Phys. Rev. B}\ }\textbf {\bibinfo {volume} {102}},\ \bibinfo
  {pages} {014313} (\bibinfo {year} {2020})}\BibitemShut {NoStop}%
\bibitem [{\citenamefont {Yao}\ \emph {et~al.}(2025)\citenamefont {Yao},
  \citenamefont {Huang}, \citenamefont {Cao}, \citenamefont {Zhang},\ and\
  \citenamefont {et.al.}}]{Yao2025}%
  \BibitemOpen
  \bibfield  {author} {\bibinfo {author} {\bibfnamefont {Y.}~\bibnamefont
  {Yao}}, \bibinfo {author} {\bibfnamefont {S.}~\bibnamefont {Huang}}, \bibinfo
  {author} {\bibfnamefont {R.}~\bibnamefont {Cao}}, \bibinfo {author}
  {\bibfnamefont {Z.}~\bibnamefont {Zhang}},\ and\ \bibinfo {author}
  {\bibnamefont {et.al.}},\ }\bibfield  {title} {\bibinfo {title}
  {{Dislocation-assisted electron and hole transport in GaN epitaxial
  layers}},\ }\href {https://doi.org/10.1038/s41467-025-61510-w} {\bibfield
  {journal} {\bibinfo  {journal} {Nature Communications}\ }\textbf {\bibinfo
  {volume} {16}},\ \bibinfo {pages} {6448} (\bibinfo {year}
  {2025})}\BibitemShut {NoStop}%
\bibitem [{\citenamefont {Reiche}\ and\ \citenamefont
  {Kittler}(2016)}]{Reiche2016}%
  \BibitemOpen
  \bibfield  {author} {\bibinfo {author} {\bibfnamefont {M.}~\bibnamefont
  {Reiche}}\ and\ \bibinfo {author} {\bibfnamefont {M.}~\bibnamefont
  {Kittler}},\ }\bibfield  {title} {\bibinfo {title} {{Electronic and Optical
  Properties of Dislocations in Silicon}},\ }\bibfield  {journal} {\bibinfo
  {journal} {Crystals}\ }\textbf {\bibinfo {volume} {6}},\ \href
  {https://doi.org/10.3390/cryst6070074} {10.3390/cryst6070074} (\bibinfo
  {year} {2016})\BibitemShut {NoStop}%
\bibitem [{\citenamefont {Yin}\ \emph {et~al.}(2016)\citenamefont {Yin},
  \citenamefont {Jiang}, \citenamefont {Qiao},\ and\ \citenamefont
  {He}}]{Yin2016}%
  \BibitemOpen
  \bibfield  {author} {\bibinfo {author} {\bibfnamefont {L.-J.}\ \bibnamefont
  {Yin}}, \bibinfo {author} {\bibfnamefont {H.}~\bibnamefont {Jiang}}, \bibinfo
  {author} {\bibfnamefont {J.-B.}\ \bibnamefont {Qiao}},\ and\ \bibinfo
  {author} {\bibfnamefont {L.}~\bibnamefont {He}},\ }\bibfield  {title}
  {\bibinfo {title} {Direct imaging of topological edge states at a bilayer
  graphene domain wall},\ }\href {https://doi.org/10.1038/ncomms11760}
  {\bibfield  {journal} {\bibinfo  {journal} {Nature Communications}\ }\textbf
  {\bibinfo {volume} {7}},\ \bibinfo {pages} {11760} (\bibinfo {year}
  {2016})}\BibitemShut {NoStop}%
\bibitem [{\citenamefont {Thiel}\ \emph {et~al.}(2009)\citenamefont {Thiel},
  \citenamefont {Schneider}, \citenamefont {Kourkoutis}, \citenamefont
  {Muller}, \citenamefont {Reyren}, \citenamefont {Caviglia}, \citenamefont
  {Gariglio}, \citenamefont {Triscone},\ and\ \citenamefont
  {Mannhart}}]{Thiel2009}%
  \BibitemOpen
  \bibfield  {author} {\bibinfo {author} {\bibfnamefont {S.}~\bibnamefont
  {Thiel}}, \bibinfo {author} {\bibfnamefont {C.~W.}\ \bibnamefont
  {Schneider}}, \bibinfo {author} {\bibfnamefont {L.~F.}\ \bibnamefont
  {Kourkoutis}}, \bibinfo {author} {\bibfnamefont {D.~A.}\ \bibnamefont
  {Muller}}, \bibinfo {author} {\bibfnamefont {N.}~\bibnamefont {Reyren}},
  \bibinfo {author} {\bibfnamefont {A.~D.}\ \bibnamefont {Caviglia}}, \bibinfo
  {author} {\bibfnamefont {S.}~\bibnamefont {Gariglio}}, \bibinfo {author}
  {\bibfnamefont {J.-M.}\ \bibnamefont {Triscone}},\ and\ \bibinfo {author}
  {\bibfnamefont {J.}~\bibnamefont {Mannhart}},\ }\bibfield  {title} {\bibinfo
  {title} {{Electron Scattering at Dislocations in
  ${\mathrm{LaAlO}}_{3}/{\mathrm{SrTiO}}_{3}$ Interfaces}},\ }\href
  {https://doi.org/10.1103/PhysRevLett.102.046809} {\bibfield  {journal}
  {\bibinfo  {journal} {Phys. Rev. Lett.}\ }\textbf {\bibinfo {volume} {102}},\
  \bibinfo {pages} {046809} (\bibinfo {year} {2009})}\BibitemShut {NoStop}%
\bibitem [{\citenamefont {Xue}\ \emph {et~al.}(2021)\citenamefont {Xue},
  \citenamefont {Jia}, \citenamefont {Ge}, \citenamefont {Guan}, \citenamefont
  {Wang}, \citenamefont {Yuan}, \citenamefont {Sun}, \citenamefont {Chong},\
  and\ \citenamefont {Zhang}}]{Xue2021}%
  \BibitemOpen
  \bibfield  {author} {\bibinfo {author} {\bibfnamefont {H.}~\bibnamefont
  {Xue}}, \bibinfo {author} {\bibfnamefont {D.}~\bibnamefont {Jia}}, \bibinfo
  {author} {\bibfnamefont {Y.}~\bibnamefont {Ge}}, \bibinfo {author}
  {\bibfnamefont {Y.-j.}\ \bibnamefont {Guan}}, \bibinfo {author}
  {\bibfnamefont {Q.}~\bibnamefont {Wang}}, \bibinfo {author} {\bibfnamefont
  {S.-q.}\ \bibnamefont {Yuan}}, \bibinfo {author} {\bibfnamefont {H.-x.}\
  \bibnamefont {Sun}}, \bibinfo {author} {\bibfnamefont {Y.~D.}\ \bibnamefont
  {Chong}},\ and\ \bibinfo {author} {\bibfnamefont {B.}~\bibnamefont {Zhang}},\
  }\bibfield  {title} {\bibinfo {title} {{Observation of Dislocation-Induced
  Topological Modes in a Three-Dimensional Acoustic Topological Insulator}},\
  }\href {https://doi.org/10.1103/PhysRevLett.127.214301} {\bibfield  {journal}
  {\bibinfo  {journal} {Phys. Rev. Lett.}\ }\textbf {\bibinfo {volume} {127}},\
  \bibinfo {pages} {214301} (\bibinfo {year} {2021})}\BibitemShut {NoStop}%
\bibitem [{\citenamefont {Bergmann}(1984)}]{Bergmann1984}%
  \BibitemOpen
  \bibfield  {author} {\bibinfo {author} {\bibfnamefont {G.}~\bibnamefont
  {Bergmann}},\ }\bibfield  {title} {\bibinfo {title} {Weak localization in
  thin films: a time-of-flight experiment with conduction electrons},\ }\href
  {https://doi.org/https://doi.org/10.1016/0370-1573(84)90103-0} {\bibfield
  {journal} {\bibinfo  {journal} {Physics Reports}\ }\textbf {\bibinfo {volume}
  {107}},\ \bibinfo {pages} {1} (\bibinfo {year} {1984})}\BibitemShut {NoStop}%
\bibitem [{\citenamefont {Altland}\ \emph {et~al.}(2002)\citenamefont
  {Altland}, \citenamefont {Simons},\ and\ \citenamefont
  {Zirnbauer}}]{Altland2002}%
  \BibitemOpen
  \bibfield  {author} {\bibinfo {author} {\bibfnamefont {A.}~\bibnamefont
  {Altland}}, \bibinfo {author} {\bibfnamefont {B.}~\bibnamefont {Simons}},\
  and\ \bibinfo {author} {\bibfnamefont {M.}~\bibnamefont {Zirnbauer}},\
  }\bibfield  {title} {\bibinfo {title} {Theories of low-energy quasi-particle
  states in disordered d-wave superconductors},\ }\href
  {https://doi.org/https://doi.org/10.1016/S0370-1573(01)00065-5} {\bibfield
  {journal} {\bibinfo  {journal} {Physics Reports}\ }\textbf {\bibinfo {volume}
  {359}},\ \bibinfo {pages} {283} (\bibinfo {year} {2002})}\BibitemShut
  {NoStop}%
\bibitem [{\citenamefont {Altshuler}\ and\ \citenamefont
  {Aronov}(1979)}]{Altshuler1979_1}%
  \BibitemOpen
  \bibfield  {author} {\bibinfo {author} {\bibfnamefont {B.~L.}\ \bibnamefont
  {Altshuler}}\ and\ \bibinfo {author} {\bibfnamefont {A.}~\bibnamefont
  {Aronov}},\ }\bibfield  {title} {\bibinfo {title} {Contribution to the theory
  of disordered metals in strongly doped semiconductors},\ }\href@noop {}
  {\bibfield  {journal} {\bibinfo  {journal} {Journal of Experimental and
  Theoretical Physics}\ }\textbf {\bibinfo {volume} {50}} (\bibinfo {year}
  {1979})}\BibitemShut {NoStop}%
\bibitem [{\citenamefont {Altshuler}\ \emph {et~al.}(1980)\citenamefont
  {Altshuler}, \citenamefont {Aronov},\ and\ \citenamefont
  {Lee}}]{Altshuler1980}%
  \BibitemOpen
  \bibfield  {author} {\bibinfo {author} {\bibfnamefont {B.~L.}\ \bibnamefont
  {Altshuler}}, \bibinfo {author} {\bibfnamefont {A.~G.}\ \bibnamefont
  {Aronov}},\ and\ \bibinfo {author} {\bibfnamefont {P.~A.}\ \bibnamefont
  {Lee}},\ }\bibfield  {title} {\bibinfo {title} {{Interaction Effects in
  Disordered Fermi Systems in Two Dimensions}},\ }\href
  {https://doi.org/10.1103/PhysRevLett.44.1288} {\bibfield  {journal} {\bibinfo
   {journal} {Phys. Rev. Lett.}\ }\textbf {\bibinfo {volume} {44}},\ \bibinfo
  {pages} {1288} (\bibinfo {year} {1980})}\BibitemShut {NoStop}%
\bibitem [{\citenamefont {Altshuler}\ and\ \citenamefont
  {Aronov}(1985)}]{Altshuler1985}%
  \BibitemOpen
  \bibfield  {author} {\bibinfo {author} {\bibfnamefont {B.}~\bibnamefont
  {Altshuler}}\ and\ \bibinfo {author} {\bibfnamefont {A.}~\bibnamefont
  {Aronov}},\ }\bibfield  {title} {\bibinfo {title} {{Chapter 1 -
  Electron–Electron Interaction In Disordered Conductors}},\ }in\ \href
  {https://doi.org/https://doi.org/10.1016/B978-0-444-86916-6.50007-7} {\emph
  {\bibinfo {booktitle} {Electron–Electron Interactions in Disordered
  Systems}}},\ \bibinfo {series} {Modern Problems in Condensed Matter
  Sciences}, Vol.~\bibinfo {volume} {10},\ \bibinfo {editor} {edited by\
  \bibinfo {editor} {\bibfnamefont {A.}~\bibnamefont {Efros}}\ and\ \bibinfo
  {editor} {\bibfnamefont {M.}~\bibnamefont {Pollak}}}\ (\bibinfo  {publisher}
  {Elsevier},\ \bibinfo {year} {1985})\ pp.\ \bibinfo {pages}
  {1--153}\BibitemShut {NoStop}%
\bibitem [{\citenamefont {{Finkel'shtein}}(1983)}]{Finkelshtein1983}%
  \BibitemOpen
  \bibfield  {author} {\bibinfo {author} {\bibfnamefont {A.~M.}\ \bibnamefont
  {{Finkel'shtein}}},\ }\bibfield  {title} {\bibinfo {title} {{{Influence of
  Coulomb interaction on the properties of disordered metals}}},\ }\href@noop
  {} {\bibfield  {journal} {\bibinfo  {journal} {Soviet Journal of Experimental
  and Theoretical Physics}\ }\textbf {\bibinfo {volume} {57}},\ \bibinfo
  {pages} {97} (\bibinfo {year} {1983})}\BibitemShut {NoStop}%
\bibitem [{\citenamefont {Fukuyama}(1985)}]{Fukuyama1985}%
  \BibitemOpen
  \bibfield  {author} {\bibinfo {author} {\bibfnamefont {H.}~\bibnamefont
  {Fukuyama}},\ }\bibfield  {title} {\bibinfo {title} {Chapter 2 - interaction
  effects in the weakly localized regime of two- and three-dimensional
  disordered systems},\ }in\ \href
  {https://doi.org/https://doi.org/10.1016/B978-0-444-86916-6.50008-9} {\emph
  {\bibinfo {booktitle} {Electron–Electron Interactions in Disordered
  Systems}}},\ \bibinfo {series} {Modern Problems in Condensed Matter
  Sciences}, Vol.~\bibinfo {volume} {10},\ \bibinfo {editor} {edited by\
  \bibinfo {editor} {\bibfnamefont {A.}~\bibnamefont {Efros}}\ and\ \bibinfo
  {editor} {\bibfnamefont {M.}~\bibnamefont {Pollak}}}\ (\bibinfo  {publisher}
  {Elsevier},\ \bibinfo {year} {1985})\ pp.\ \bibinfo {pages}
  {155--230}\BibitemShut {NoStop}%
\bibitem [{\citenamefont {Firsov}\ \emph {et~al.}(1987)\citenamefont {Firsov},
  \citenamefont {Nakhmebov}, \citenamefont {Prigodin},\ and\ \citenamefont
  {Weller}}]{Firsov1987}%
  \BibitemOpen
  \bibfield  {author} {\bibinfo {author} {\bibfnamefont {Y.~A.}\ \bibnamefont
  {Firsov}}, \bibinfo {author} {\bibfnamefont {E.~P.}\ \bibnamefont
  {Nakhmebov}}, \bibinfo {author} {\bibfnamefont {V.~N.}\ \bibnamefont
  {Prigodin}},\ and\ \bibinfo {author} {\bibfnamefont {W.}~\bibnamefont
  {Weller}},\ }\bibfield  {title} {\bibinfo {title} {{Corrections from
  Electron-Electron Interaction to the One-Particle Density of States in
  Strongly Anisotropic Two-Dimensional Systems}},\ }\href
  {https://doi.org/https://doi.org/10.1002/pssb.2221410110} {\bibfield
  {journal} {\bibinfo  {journal} {physica status solidi (b)}\ }\textbf
  {\bibinfo {volume} {141}},\ \bibinfo {pages} {111} (\bibinfo {year}
  {1987})}\BibitemShut {NoStop}%
\bibitem [{\citenamefont {Firsov}\ \emph {et~al.}(1988)\citenamefont {Firsov},
  \citenamefont {Nakhmedov}, \citenamefont {Prigodin},\ and\ \citenamefont
  {Weller}}]{Firsov1988}%
  \BibitemOpen
  \bibfield  {author} {\bibinfo {author} {\bibfnamefont {Y.~A.}\ \bibnamefont
  {Firsov}}, \bibinfo {author} {\bibfnamefont {E.~P.}\ \bibnamefont
  {Nakhmedov}}, \bibinfo {author} {\bibfnamefont {V.~N.}\ \bibnamefont
  {Prigodin}},\ and\ \bibinfo {author} {\bibfnamefont {W.}~\bibnamefont
  {Weller}},\ }\bibfield  {title} {\bibinfo {title} {{Low-Temperature
  Conductivity of a Weakly Disordered, Strongly Anisotropic Two-Dimensional
  System with Electron—Electron Interaction. Application to Germanium
  Bicrystals and to Surface Superlattices}},\ }\href
  {https://doi.org/https://doi.org/10.1002/pssb.2221450229} {\bibfield
  {journal} {\bibinfo  {journal} {physica status solidi (b)}\ }\textbf
  {\bibinfo {volume} {145}},\ \bibinfo {pages} {625} (\bibinfo {year}
  {1988})}\BibitemShut {NoStop}%
\bibitem [{\citenamefont {Caliskan}\ and\ \citenamefont
  {Kumru}(2012)}]{Caliskan2012}%
  \BibitemOpen
  \bibfield  {author} {\bibinfo {author} {\bibfnamefont {S.}~\bibnamefont
  {Caliskan}}\ and\ \bibinfo {author} {\bibfnamefont {M.}~\bibnamefont
  {Kumru}},\ }\bibfield  {title} {\bibinfo {title} {High-order perturbation
  corrections to the density of states of disordered metals in a magnetic
  field},\ }\href {https://doi.org/10.1103/PhysRevB.85.205148} {\bibfield
  {journal} {\bibinfo  {journal} {Phys. Rev. B}\ }\textbf {\bibinfo {volume}
  {85}},\ \bibinfo {pages} {205148} (\bibinfo {year} {2012})}\BibitemShut
  {NoStop}%
\bibitem [{\citenamefont {Zhou}\ and\ \citenamefont {Guo}(2019)}]{Zhou2019}%
  \BibitemOpen
  \bibfield  {author} {\bibinfo {author} {\bibfnamefont {C.}~\bibnamefont
  {Zhou}}\ and\ \bibinfo {author} {\bibfnamefont {H.}~\bibnamefont {Guo}},\
  }\bibfield  {title} {\bibinfo {title} {{Altshuler-Aronov effects in
  nonequilibrium disordered nanostructures}},\ }\href
  {https://doi.org/10.1103/PhysRevB.100.045413} {\bibfield  {journal} {\bibinfo
   {journal} {Phys. Rev. B}\ }\textbf {\bibinfo {volume} {100}},\ \bibinfo
  {pages} {045413} (\bibinfo {year} {2019})}\BibitemShut {NoStop}%
\bibitem [{\citenamefont {Carbillet}\ \emph {et~al.}(2020)\citenamefont
  {Carbillet}, \citenamefont {Cherkez}, \citenamefont {Skvortsov},
  \citenamefont {Feigel'man}, \citenamefont {Debontridder}, \citenamefont
  {Ioffe}, \citenamefont {Stolyarov}, \citenamefont {Ilin}, \citenamefont
  {Siegel}, \citenamefont {Roditchev}, \citenamefont {Cren},\ and\
  \citenamefont {Brun}}]{Carbillet2020}%
  \BibitemOpen
  \bibfield  {author} {\bibinfo {author} {\bibfnamefont {C.}~\bibnamefont
  {Carbillet}}, \bibinfo {author} {\bibfnamefont {V.}~\bibnamefont {Cherkez}},
  \bibinfo {author} {\bibfnamefont {M.~A.}\ \bibnamefont {Skvortsov}}, \bibinfo
  {author} {\bibfnamefont {M.~V.}\ \bibnamefont {Feigel'man}}, \bibinfo
  {author} {\bibfnamefont {F.}~\bibnamefont {Debontridder}}, \bibinfo {author}
  {\bibfnamefont {L.~B.}\ \bibnamefont {Ioffe}}, \bibinfo {author}
  {\bibfnamefont {V.~S.}\ \bibnamefont {Stolyarov}}, \bibinfo {author}
  {\bibfnamefont {K.}~\bibnamefont {Ilin}}, \bibinfo {author} {\bibfnamefont
  {M.}~\bibnamefont {Siegel}}, \bibinfo {author} {\bibfnamefont
  {D.}~\bibnamefont {Roditchev}}, \bibinfo {author} {\bibfnamefont
  {T.}~\bibnamefont {Cren}},\ and\ \bibinfo {author} {\bibfnamefont
  {C.}~\bibnamefont {Brun}},\ }\bibfield  {title} {\bibinfo {title}
  {{Spectroscopic evidence for strong correlations between local
  superconducting gap and local Altshuler-Aronov density of states suppression
  in ultrathin NbN films}},\ }\href
  {https://doi.org/10.1103/PhysRevB.102.024504} {\bibfield  {journal} {\bibinfo
   {journal} {Phys. Rev. B}\ }\textbf {\bibinfo {volume} {102}},\ \bibinfo
  {pages} {024504} (\bibinfo {year} {2020})}\BibitemShut {NoStop}%
\bibitem [{\citenamefont {Kuzmiak}\ \emph {et~al.}(2023)\citenamefont
  {Kuzmiak}, \citenamefont {Kop\ifmmode~\check{c}\else \v{c}\fi{}\'{\i}k},
  \citenamefont {Ko\ifmmode~\check{s}\else \v{s}\fi{}uth}, \citenamefont
  {Va\ifmmode~\check{n}\else \v{n}\fi{}o}, \citenamefont
  {Hani\ifmmode~\check{s}\else \v{s}\fi{}}, \citenamefont {Samuely},
  \citenamefont {Latyshev}, \citenamefont {Onufriienko}, \citenamefont
  {Komanick\'y}, \citenamefont {Ka\ifmmode \check{c}\else
  \v{c}\fi{}mar\ifmmode~\check{c}\else \v{c}\fi{}\'{\i}k}, \citenamefont
  {\ifmmode \check{Z}\else \v{Z}\fi{}emli\ifmmode~\check{c}\else \v{c}\fi{}ka},
  \citenamefont {Gmitra}, \citenamefont {Szab\'o},\ and\ \citenamefont
  {Samuely}}]{Kuzmiak2023}%
  \BibitemOpen
  \bibfield  {author} {\bibinfo {author} {\bibfnamefont {M.}~\bibnamefont
  {Kuzmiak}}, \bibinfo {author} {\bibfnamefont {M.}~\bibnamefont
  {Kop\ifmmode~\check{c}\else \v{c}\fi{}\'{\i}k}}, \bibinfo {author}
  {\bibfnamefont {F.}~\bibnamefont {Ko\ifmmode~\check{s}\else \v{s}\fi{}uth}},
  \bibinfo {author} {\bibfnamefont {V.}~\bibnamefont {Va\ifmmode~\check{n}\else
  \v{n}\fi{}o}}, \bibinfo {author} {\bibfnamefont {J.}~\bibnamefont
  {Hani\ifmmode~\check{s}\else \v{s}\fi{}}}, \bibinfo {author} {\bibfnamefont
  {T.}~\bibnamefont {Samuely}}, \bibinfo {author} {\bibfnamefont
  {V.}~\bibnamefont {Latyshev}}, \bibinfo {author} {\bibfnamefont
  {O.}~\bibnamefont {Onufriienko}}, \bibinfo {author} {\bibfnamefont
  {V.}~\bibnamefont {Komanick\'y}}, \bibinfo {author} {\bibfnamefont
  {J.}~\bibnamefont {Ka\ifmmode \check{c}\else
  \v{c}\fi{}mar\ifmmode~\check{c}\else \v{c}\fi{}\'{\i}k}}, \bibinfo {author}
  {\bibfnamefont {M.}~\bibnamefont {\ifmmode \check{Z}\else
  \v{Z}\fi{}emli\ifmmode~\check{c}\else \v{c}\fi{}ka}}, \bibinfo {author}
  {\bibfnamefont {M.}~\bibnamefont {Gmitra}}, \bibinfo {author} {\bibfnamefont
  {P.}~\bibnamefont {Szab\'o}},\ and\ \bibinfo {author} {\bibfnamefont
  {P.}~\bibnamefont {Samuely}},\ }\bibfield  {title} {\bibinfo {title}
  {{Disorder- and magnetic field--tuned fermionic superconductor-insulator
  transition in MoN thin films: Transport and scanning tunneling microscopy}},\
  }\href {https://doi.org/10.1103/PhysRevB.108.184511} {\bibfield  {journal}
  {\bibinfo  {journal} {Phys. Rev. B}\ }\textbf {\bibinfo {volume} {108}},\
  \bibinfo {pages} {184511} (\bibinfo {year} {2023})}\BibitemShut {NoStop}%
\bibitem [{\citenamefont {Liz\'ee}\ \emph {et~al.}(2023)\citenamefont
  {Liz\'ee}, \citenamefont {Stosiek}, \citenamefont {Burmistrov}, \citenamefont
  {Cren},\ and\ \citenamefont {Brun}}]{Mathieu2023}%
  \BibitemOpen
  \bibfield  {author} {\bibinfo {author} {\bibfnamefont {M.}~\bibnamefont
  {Liz\'ee}}, \bibinfo {author} {\bibfnamefont {M.}~\bibnamefont {Stosiek}},
  \bibinfo {author} {\bibfnamefont {I.}~\bibnamefont {Burmistrov}}, \bibinfo
  {author} {\bibfnamefont {T.}~\bibnamefont {Cren}},\ and\ \bibinfo {author}
  {\bibfnamefont {C.}~\bibnamefont {Brun}},\ }\bibfield  {title} {\bibinfo
  {title} {Local density of states fluctuations in a two-dimensional
  superconductor as a probe of quantum diffusion},\ }\href
  {https://doi.org/10.1103/PhysRevB.107.174508} {\bibfield  {journal} {\bibinfo
   {journal} {Phys. Rev. B}\ }\textbf {\bibinfo {volume} {107}},\ \bibinfo
  {pages} {174508} (\bibinfo {year} {2023})}\BibitemShut {NoStop}%
\bibitem [{\citenamefont {Shi}\ \emph {et~al.}(2023)\citenamefont {Shi},
  \citenamefont {Gao}, \citenamefont {Li}, \citenamefont {Zhang}, \citenamefont
  {Gornyi}, \citenamefont {Gutman},\ and\ \citenamefont {Li}}]{Shi2023}%
  \BibitemOpen
  \bibfield  {author} {\bibinfo {author} {\bibfnamefont {G.}~\bibnamefont
  {Shi}}, \bibinfo {author} {\bibfnamefont {F.}~\bibnamefont {Gao}}, \bibinfo
  {author} {\bibfnamefont {Z.}~\bibnamefont {Li}}, \bibinfo {author}
  {\bibfnamefont {R.}~\bibnamefont {Zhang}}, \bibinfo {author} {\bibfnamefont
  {I.}~\bibnamefont {Gornyi}}, \bibinfo {author} {\bibfnamefont
  {D.}~\bibnamefont {Gutman}},\ and\ \bibinfo {author} {\bibfnamefont
  {Y.}~\bibnamefont {Li}},\ }\bibfield  {title} {\bibinfo {title} {Quantum
  corrections to the magnetoconductivity of surface states in three-dimensional
  topological insulators},\ }\bibfield  {journal} {\bibinfo  {journal} {Nature
  Communications}\ }\textbf {\bibinfo {volume} {14}},\ \href
  {https://doi.org/10.1038/s41467-023-38256-4} {10.1038/s41467-023-38256-4}
  (\bibinfo {year} {2023})\BibitemShut {NoStop}%
\bibitem [{\citenamefont {Mošková}\ and\ \citenamefont
  {Moško}(2018)}]{Antonia2018}%
  \BibitemOpen
  \bibfield  {author} {\bibinfo {author} {\bibfnamefont {A.}~\bibnamefont
  {Mošková}}\ and\ \bibinfo {author} {\bibfnamefont {M.}~\bibnamefont
  {Moško}},\ }\bibfield  {title} {\bibinfo {title} {{States-conserving density
  of states for Altshuler-Aronov effect: Heuristic derivation}},\ }\href
  {https://doi.org/https://doi.org/10.1016/j.ssc.2018.08.010} {\bibfield
  {journal} {\bibinfo  {journal} {Solid State Communications}\ }\textbf
  {\bibinfo {volume} {284-286}},\ \bibinfo {pages} {56} (\bibinfo {year}
  {2018})}\BibitemShut {NoStop}%
\bibitem [{\citenamefont {Huang}\ and\ \citenamefont
  {Das~Sarma}(2024)}]{Huang2024}%
  \BibitemOpen
  \bibfield  {author} {\bibinfo {author} {\bibfnamefont {Y.}~\bibnamefont
  {Huang}}\ and\ \bibinfo {author} {\bibfnamefont {S.}~\bibnamefont
  {Das~Sarma}},\ }\bibfield  {title} {\bibinfo {title} {{Electronic transport,
  metal-insulator transition, and Wigner crystallization in transition metal
  dichalcogenide monolayers}},\ }\href
  {https://doi.org/10.1103/PhysRevB.109.245431} {\bibfield  {journal} {\bibinfo
   {journal} {Phys. Rev. B}\ }\textbf {\bibinfo {volume} {109}},\ \bibinfo
  {pages} {245431} (\bibinfo {year} {2024})}\BibitemShut {NoStop}%
\bibitem [{\citenamefont {Rollb\"uhler}\ and\ \citenamefont
  {Grabert}(2003)}]{Jorg2003}%
  \BibitemOpen
  \bibfield  {author} {\bibinfo {author} {\bibfnamefont {J.}~\bibnamefont
  {Rollb\"uhler}}\ and\ \bibinfo {author} {\bibfnamefont {H.}~\bibnamefont
  {Grabert}},\ }\bibfield  {title} {\bibinfo {title} {{Tunneling Density of
  States of the Interacting Two-Dimensional Electron Gas}},\ }\href
  {https://doi.org/10.1103/PhysRevLett.91.166402} {\bibfield  {journal}
  {\bibinfo  {journal} {Phys. Rev. Lett.}\ }\textbf {\bibinfo {volume} {91}},\
  \bibinfo {pages} {166402} (\bibinfo {year} {2003})}\BibitemShut {NoStop}%
\bibitem [{\citenamefont {Romano}\ \emph {et~al.}(2005)\citenamefont {Romano},
  \citenamefont {Ulloa},\ and\ \citenamefont {Tamborenea}}]{Romano2005}%
  \BibitemOpen
  \bibfield  {author} {\bibinfo {author} {\bibfnamefont {C.~L.}\ \bibnamefont
  {Romano}}, \bibinfo {author} {\bibfnamefont {S.~E.}\ \bibnamefont {Ulloa}},\
  and\ \bibinfo {author} {\bibfnamefont {P.~I.}\ \bibnamefont {Tamborenea}},\
  }\bibfield  {title} {\bibinfo {title} {Level structure and spin-orbit effects
  in quasi-one-dimensional semiconductor nanostructures},\ }\href
  {https://doi.org/10.1103/PhysRevB.71.035336} {\bibfield  {journal} {\bibinfo
  {journal} {Phys. Rev. B}\ }\textbf {\bibinfo {volume} {71}},\ \bibinfo
  {pages} {035336} (\bibinfo {year} {2005})}\BibitemShut {NoStop}%
\bibitem [{\citenamefont {Suleymanli}\ \emph {et~al.}(2023)\citenamefont
  {Suleymanli}, \citenamefont {Nakhmedov}, \citenamefont {Tatardar},\ and\
  \citenamefont {Tanatar}}]{Suleymanli2023}%
  \BibitemOpen
  \bibfield  {author} {\bibinfo {author} {\bibfnamefont {B.}~\bibnamefont
  {Suleymanli}}, \bibinfo {author} {\bibfnamefont {E.}~\bibnamefont
  {Nakhmedov}}, \bibinfo {author} {\bibfnamefont {F.}~\bibnamefont
  {Tatardar}},\ and\ \bibinfo {author} {\bibfnamefont {B.}~\bibnamefont
  {Tanatar}},\ }\bibfield  {title} {\bibinfo {title} {{The diagrammatic method
  of Berezinskii for one-dimensional disordered wire with spin–orbit
  interaction}},\ }\href
  {https://doi.org/https://doi.org/10.1016/j.physe.2022.115550} {\bibfield
  {journal} {\bibinfo  {journal} {Physica E: Low-dimensional Systems and
  Nanostructures}\ }\textbf {\bibinfo {volume} {146}},\ \bibinfo {pages}
  {115550} (\bibinfo {year} {2023})}\BibitemShut {NoStop}%
\bibitem [{\citenamefont {Kane}\ \emph {et~al.}(2002)\citenamefont {Kane},
  \citenamefont {Mukhopadhyay},\ and\ \citenamefont {Lubensky}}]{Kane2002}%
  \BibitemOpen
  \bibfield  {author} {\bibinfo {author} {\bibfnamefont {C.~L.}\ \bibnamefont
  {Kane}}, \bibinfo {author} {\bibfnamefont {R.}~\bibnamefont {Mukhopadhyay}},\
  and\ \bibinfo {author} {\bibfnamefont {T.~C.}\ \bibnamefont {Lubensky}},\
  }\bibfield  {title} {\bibinfo {title} {Fractional quantum hall effect in an
  array of quantum wires},\ }\href
  {https://doi.org/10.1103/PhysRevLett.88.036401} {\bibfield  {journal}
  {\bibinfo  {journal} {Phys. Rev. Lett.}\ }\textbf {\bibinfo {volume} {88}},\
  \bibinfo {pages} {036401} (\bibinfo {year} {2002})}\BibitemShut {NoStop}%
\bibitem [{\citenamefont {Klinovaja}\ and\ \citenamefont
  {Loss}(2013)}]{Klinovaja2013}%
  \BibitemOpen
  \bibfield  {author} {\bibinfo {author} {\bibfnamefont {J.}~\bibnamefont
  {Klinovaja}}\ and\ \bibinfo {author} {\bibfnamefont {D.}~\bibnamefont
  {Loss}},\ }\bibfield  {title} {\bibinfo {title} {Topological edge states and
  fractional quantum hall effect from umklapp scattering},\ }\href
  {https://doi.org/10.1103/PhysRevLett.111.196401} {\bibfield  {journal}
  {\bibinfo  {journal} {Phys. Rev. Lett.}\ }\textbf {\bibinfo {volume} {111}},\
  \bibinfo {pages} {196401} (\bibinfo {year} {2013})}\BibitemShut {NoStop}%
\bibitem [{\citenamefont {Burkov}\ \emph {et~al.}(2004)\citenamefont {Burkov},
  \citenamefont {N\'u\~nez},\ and\ \citenamefont {MacDonald}}]{Burkov2004}%
  \BibitemOpen
  \bibfield  {author} {\bibinfo {author} {\bibfnamefont {A.~A.}\ \bibnamefont
  {Burkov}}, \bibinfo {author} {\bibfnamefont {A.~S.}\ \bibnamefont
  {N\'u\~nez}},\ and\ \bibinfo {author} {\bibfnamefont {A.~H.}\ \bibnamefont
  {MacDonald}},\ }\bibfield  {title} {\bibinfo {title} {{Theory of
  spin-charge-coupled transport in a two-dimensional electron gas with Rashba
  spin-orbit interactions}},\ }\href
  {https://doi.org/10.1103/PhysRevB.70.155308} {\bibfield  {journal} {\bibinfo
  {journal} {Phys. Rev. B}\ }\textbf {\bibinfo {volume} {70}},\ \bibinfo
  {pages} {155308} (\bibinfo {year} {2004})}\BibitemShut {NoStop}%
\bibitem [{\citenamefont {Abrikosov}\ \emph {et~al.}(2012)\citenamefont
  {Abrikosov}, \citenamefont {Gorkov},\ and\ \citenamefont
  {Dzyaloshinski}}]{Abrikosov2012}%
  \BibitemOpen
  \bibfield  {author} {\bibinfo {author} {\bibfnamefont {A.}~\bibnamefont
  {Abrikosov}}, \bibinfo {author} {\bibfnamefont {L.}~\bibnamefont {Gorkov}},\
  and\ \bibinfo {author} {\bibfnamefont {I.}~\bibnamefont {Dzyaloshinski}},\
  }\href@noop {} {\emph {\bibinfo {title} {{Methods of Quantum Field Theory in
  Statistical Physics}}}},\ Dover Books on Physics\ (\bibinfo  {publisher}
  {Dover Publications},\ \bibinfo {year} {2012})\BibitemShut {NoStop}%
\bibitem [{\citenamefont {Arfken}\ \emph {et~al.}(2013)\citenamefont {Arfken},
  \citenamefont {Weber},\ and\ \citenamefont {Harris}}]{Arfken2013}%
  \BibitemOpen
  \bibfield  {author} {\bibinfo {author} {\bibfnamefont {G.~B.}\ \bibnamefont
  {Arfken}}, \bibinfo {author} {\bibfnamefont {H.~J.}\ \bibnamefont {Weber}},\
  and\ \bibinfo {author} {\bibfnamefont {F.~E.}\ \bibnamefont {Harris}},\
  }\bibfield  {title} {\bibinfo {title} {{Chapter 11 - Complex Variable
  Theory}},\ }in\ \href
  {https://doi.org/https://doi.org/10.1016/B978-0-12-384654-9.00011-6} {\emph
  {\bibinfo {booktitle} {{Mathematical Methods for Physicists (Seventh
  Edition)}}}},\ \bibinfo {editor} {edited by\ \bibinfo {editor} {\bibfnamefont
  {G.~B.}\ \bibnamefont {Arfken}}, \bibinfo {editor} {\bibfnamefont {H.~J.}\
  \bibnamefont {Weber}},\ and\ \bibinfo {editor} {\bibfnamefont {F.~E.}\
  \bibnamefont {Harris}}}\ (\bibinfo  {publisher} {Academic Press},\ \bibinfo
  {address} {Boston},\ \bibinfo {year} {2013})\ \bibinfo {edition} {seventh
  edition}\ ed.,\ pp.\ \bibinfo {pages} {469--550}\BibitemShut {NoStop}%
\bibitem [{\citenamefont {Hamaguchi}(2009)}]{Hamaguchi2009}%
  \BibitemOpen
  \bibfield  {author} {\bibinfo {author} {\bibfnamefont {C.}~\bibnamefont
  {Hamaguchi}},\ }\href@noop {} {\emph {\bibinfo {title} {{Basic Semiconductor
  Physics}}}}\ (\bibinfo  {publisher} {Springer Berlin Heidelberg},\ \bibinfo
  {year} {2009})\BibitemShut {NoStop}%
\bibitem [{\citenamefont {Altland}\ and\ \citenamefont
  {Simons}(2010)}]{Altland2010}%
  \BibitemOpen
  \bibfield  {author} {\bibinfo {author} {\bibfnamefont {A.}~\bibnamefont
  {Altland}}\ and\ \bibinfo {author} {\bibfnamefont {B.}~\bibnamefont
  {Simons}},\ }\href@noop {} {\emph {\bibinfo {title} {Condensed Matter Field
  Theory}}},\ Cambridge books online\ (\bibinfo  {publisher} {Cambridge
  University Press},\ \bibinfo {year} {2010})\BibitemShut {NoStop}%
\bibitem [{\citenamefont {Şaşıoğlu}\ \emph {et~al.}(2009)\citenamefont
  {Şaşıoğlu}, \citenamefont {Çalışkan},\ and\ \citenamefont
  {Kumru}}]{Sasioglu2009}%
  \BibitemOpen
  \bibfield  {author} {\bibinfo {author} {\bibfnamefont {E.}~\bibnamefont
  {Şaşıoğlu}}, \bibinfo {author} {\bibfnamefont {S.}~\bibnamefont
  {Çalışkan}},\ and\ \bibinfo {author} {\bibfnamefont {M.}~\bibnamefont
  {Kumru}},\ }\bibfield  {title} {\bibinfo {title} {{Critical behavior of
  density of states near Fermi energy in low-dimensional disordered metals}},\
  }\href {https://doi.org/10.1103/PhysRevB.79.035123} {\bibfield  {journal}
  {\bibinfo  {journal} {Phys. Rev. B}\ }\textbf {\bibinfo {volume} {79}},\
  \bibinfo {pages} {035123} (\bibinfo {year} {2009})}\BibitemShut {NoStop}%
\bibitem [{\citenamefont {Niu}\ \emph {et~al.}(2020)\citenamefont {Niu},
  \citenamefont {Qiu}, \citenamefont {Wang}, \citenamefont {Zhang},
  \citenamefont {Si}, \citenamefont {Wu},\ and\ \citenamefont {Ye}}]{Niu2020}%
  \BibitemOpen
  \bibfield  {author} {\bibinfo {author} {\bibfnamefont {C.}~\bibnamefont
  {Niu}}, \bibinfo {author} {\bibfnamefont {G.}~\bibnamefont {Qiu}}, \bibinfo
  {author} {\bibfnamefont {Y.}~\bibnamefont {Wang}}, \bibinfo {author}
  {\bibfnamefont {Z.}~\bibnamefont {Zhang}}, \bibinfo {author} {\bibfnamefont
  {M.}~\bibnamefont {Si}}, \bibinfo {author} {\bibfnamefont {W.}~\bibnamefont
  {Wu}},\ and\ \bibinfo {author} {\bibfnamefont {P.~D.}\ \bibnamefont {Ye}},\
  }\bibfield  {title} {\bibinfo {title} {Gate-tunable strong spin-orbit
  interaction in two-dimensional tellurium probed by weak antilocalization},\
  }\href {https://doi.org/10.1103/PhysRevB.101.205414} {\bibfield  {journal}
  {\bibinfo  {journal} {Phys. Rev. B}\ }\textbf {\bibinfo {volume} {101}},\
  \bibinfo {pages} {205414} (\bibinfo {year} {2020})}\BibitemShut {NoStop}%
\bibitem [{\citenamefont {Geldiyev}\ \emph {et~al.}(2023)\citenamefont
  {Geldiyev}, \citenamefont {\"Unzelmann}, \citenamefont {Eck}, \citenamefont
  {Ki\ss{}linger}, \citenamefont {Schusser}, \citenamefont {Figgemeier},
  \citenamefont {Kagerer}, \citenamefont {Tezak}, \citenamefont {Krivenkov},
  \citenamefont {Varykhalov}, \citenamefont {Fedorov}, \citenamefont
  {Nicola\"{\i}}, \citenamefont {Min\'ar}, \citenamefont {Miyamoto},
  \citenamefont {Okuda}, \citenamefont {Shimada}, \citenamefont {Di~Sante},
  \citenamefont {Sangiovanni}, \citenamefont {Hammer}, \citenamefont
  {Schneider}, \citenamefont {Bentmann},\ and\ \citenamefont
  {Reinert}}]{Geldiyev2023}%
  \BibitemOpen
  \bibfield  {author} {\bibinfo {author} {\bibfnamefont {B.}~\bibnamefont
  {Geldiyev}}, \bibinfo {author} {\bibfnamefont {M.}~\bibnamefont
  {\"Unzelmann}}, \bibinfo {author} {\bibfnamefont {P.}~\bibnamefont {Eck}},
  \bibinfo {author} {\bibfnamefont {T.}~\bibnamefont {Ki\ss{}linger}}, \bibinfo
  {author} {\bibfnamefont {J.}~\bibnamefont {Schusser}}, \bibinfo {author}
  {\bibfnamefont {T.}~\bibnamefont {Figgemeier}}, \bibinfo {author}
  {\bibfnamefont {P.}~\bibnamefont {Kagerer}}, \bibinfo {author} {\bibfnamefont
  {N.}~\bibnamefont {Tezak}}, \bibinfo {author} {\bibfnamefont
  {M.}~\bibnamefont {Krivenkov}}, \bibinfo {author} {\bibfnamefont
  {A.}~\bibnamefont {Varykhalov}}, \bibinfo {author} {\bibfnamefont
  {A.}~\bibnamefont {Fedorov}}, \bibinfo {author} {\bibfnamefont
  {L.}~\bibnamefont {Nicola\"{\i}}}, \bibinfo {author} {\bibfnamefont
  {J.}~\bibnamefont {Min\'ar}}, \bibinfo {author} {\bibfnamefont
  {K.}~\bibnamefont {Miyamoto}}, \bibinfo {author} {\bibfnamefont
  {T.}~\bibnamefont {Okuda}}, \bibinfo {author} {\bibfnamefont
  {K.}~\bibnamefont {Shimada}}, \bibinfo {author} {\bibfnamefont
  {D.}~\bibnamefont {Di~Sante}}, \bibinfo {author} {\bibfnamefont
  {G.}~\bibnamefont {Sangiovanni}}, \bibinfo {author} {\bibfnamefont
  {L.}~\bibnamefont {Hammer}}, \bibinfo {author} {\bibfnamefont {M.~A.}\
  \bibnamefont {Schneider}}, \bibinfo {author} {\bibfnamefont {H.}~\bibnamefont
  {Bentmann}},\ and\ \bibinfo {author} {\bibfnamefont {F.}~\bibnamefont
  {Reinert}},\ }\bibfield  {title} {\bibinfo {title} {Strongly anisotropic spin
  and orbital rashba effect at a tellurium -- noble metal interface},\ }\href
  {https://doi.org/10.1103/PhysRevB.108.L121107} {\bibfield  {journal}
  {\bibinfo  {journal} {Phys. Rev. B}\ }\textbf {\bibinfo {volume} {108}},\
  \bibinfo {pages} {L121107} (\bibinfo {year} {2023})}\BibitemShut {NoStop}%
\bibitem [{\citenamefont {Liu}\ \emph {et~al.}(2014)\citenamefont {Liu},
  \citenamefont {Niu}, \citenamefont {Xiang}, \citenamefont {Wei},
  \citenamefont {Li}, \citenamefont {Feng}, \citenamefont {Han}, \citenamefont
  {Zhang},\ and\ \citenamefont {Coey}}]{Liu2014}%
  \BibitemOpen
  \bibfield  {author} {\bibinfo {author} {\bibfnamefont {L.}~\bibnamefont
  {Liu}}, \bibinfo {author} {\bibfnamefont {J.}~\bibnamefont {Niu}}, \bibinfo
  {author} {\bibfnamefont {L.}~\bibnamefont {Xiang}}, \bibinfo {author}
  {\bibfnamefont {J.}~\bibnamefont {Wei}}, \bibinfo {author} {\bibfnamefont
  {D.-L.}\ \bibnamefont {Li}}, \bibinfo {author} {\bibfnamefont {J.-F.}\
  \bibnamefont {Feng}}, \bibinfo {author} {\bibfnamefont {X.-F.}\ \bibnamefont
  {Han}}, \bibinfo {author} {\bibfnamefont {X.-G.}\ \bibnamefont {Zhang}},\
  and\ \bibinfo {author} {\bibfnamefont {J.~M.~D.}\ \bibnamefont {Coey}},\
  }\bibfield  {title} {\bibinfo {title} {Symmetry-dependent electron-electron
  interaction in coherent tunnel junctions resolved by measurements of
  zero-bias anomaly},\ }\href {https://doi.org/10.1103/PhysRevB.90.195132}
  {\bibfield  {journal} {\bibinfo  {journal} {Phys. Rev. B}\ }\textbf {\bibinfo
  {volume} {90}},\ \bibinfo {pages} {195132} (\bibinfo {year}
  {2014})}\BibitemShut {NoStop}%
\end{thebibliography}%

\end{document}